\documentclass[aps,prb,10pt,twocolumn,superscriptaddress,nofootinbib,floatfix]{revtex4-2}
\usepackage{amsmath,amssymb,bm}
\usepackage{graphicx}
\usepackage{booktabs}
\usepackage{placeins}
\usepackage{array}
\usepackage{xcolor}
\usepackage{hyperref}
\usepackage{orcidlink}
\hypersetup{hidelinks}
\graphicspath{{./}{figures/}}
\begin{document}
\title{Finite-Time Thermodynamic Routing in an Anharmonic Collective Quantum Otto Engine}
\author{Gabriella G. Damas~\orcidlink{0000-0003-3376-9281}}
\affiliation{Instituto de F\'isica, Universidade Federal de Goi\'as, 74.001-970, Goi\^ania--GO, Brazil}

\author{G. D. de Moraes Neto~\orcidlink{0000-0003-4273-8380}}
\email{gentil.neto@hibiuh.edu.cn}
\affiliation{Department of Fundamental Sciences, Hainan Bielefeld University of Applied Sciences, Danzhou, Hainan 578101, China}
\begin{abstract}
The Gibbs state determined by a thermal reservoir does not determine how a working medium approaches it or how fluctuations accumulate over repeated cycles. We exploit this kinetic freedom in an anharmonic collective quantum Otto engine. A longitudinal interaction separates neighboring Dicke transition frequencies without changing their collective matrix elements, allowing passive spectral shaping to route finite-time transport among edges with different heat costs while preserving the contact Gibbs states. Exact edge-current identities and repeated-cycle counting statistics link this routing to mean work, hot heat, and intercycle work correlations. At fixed thermal-contact exposure, structured reservoirs can increase both mean work and long-time squared precision; equal-work controls distinguish these gains from uniform relaxation-rate changes. The useful regime is mesoscopic, bounded by spectral resolution and physical contact time. An equal-cooling example and a full-spin local-noise study test the mechanism beyond ideal engine operation, the latter retaining a one-cycle precision gain despite spin-sector mixing. In a six-transmon dressed-spectrum model, passive microwave filters at fixed integrated coupling achieve equal mean work with shorter thermal contacts, corresponding to higher contact-time-normalized output, and with long-time squared precision comparable to that of the fastest sampled flat reference. The transition pathways selected by a passive reservoir thus provide a control variable for finite-time performance and accumulated work fluctuations that is independent of thermal equilibrium.
\end{abstract}
\maketitle

\section{Introduction}
\label{sec:introduction}

A thermal reservoir determines not only the equilibrium state of a quantum working medium but also the transition pathways through which that state is approached. Detailed balance fixes the ratio of upward and downward transition rates, while the reservoir spectrum can redistribute finite-time energy transport among those transitions. The quantum Otto cycle provides a natural setting to examine this effect because its isolated work strokes and dissipative thermal contacts are separated by construction~\cite{Quan2007,TalknerLutzHanggi2007,EspositoHarbolaMukamel2009}. For microscopic engines, the consequences extend beyond average work extraction. Work and heat fluctuate from cycle to cycle, incomplete thermalization can correlate successive cycles, and experiments can resolve joint work--heat statistics~\cite{DenzlerLutz2020,SaryalAgarwalla2021,Bouton2021,DenzlerEtAl2024}. Finite-time performance requires characterizing both useful output and its statistical precision.

Collective working media make these kinetic effects particularly rich. When many constituents exchange energy through a common transition operator, their amplitudes combine into Dicke matrix elements, producing strongly excitation-dependent transition rates~\cite{Dicke1954,GrossHaroche1982}. Collective coupling has consequently been used to accelerate thermalization and improve the power, stability, and reliability of quantum thermal machines~\cite{HardalMustecaplioglu2015,NiedenzuKurizki2018,Kloc2019,Souza2022,Jaseem2023}. Related Otto machines have also been studied using the open Dicke model~\cite{Xu2024}. These developments establish the role of cooperative dynamics, but collective enhancement alone does not provide independent control over the different transitions of an equally spaced ladder. When the ladder is anharmonic, neighboring transitions occur at distinct Bohr frequencies. A structured reservoir can then assign different kinetic weights to different rungs while retaining their collective matrix-element enhancement.

Reservoir engineering and frequency-selective collective absorption have demonstrated the opportunities offered by transition-dependent dissipation~\cite{GelbwaserKlimovsky2013,Higgins2014,Naseem2020,Kamimura2022}. Recent developments include finite-bandwidth filtering of parasitic heat-transfer channels, where spectral selectivity competes with the finite response time of a physical filter~\cite{Prataviera2026}, and experimental control of microscopic relaxation rates to optimize the finite-time power of an atomic Otto engine~\cite{Burgardt2026}. Finite-contact work statistics and correlations have also been analyzed for collective Dicke Otto engines~\cite{GarcezEtAlCollective2026}. Here the distinct control is passive and frequency selective: the reservoir redistributes transport among thermodynamically inequivalent, collectively enhanced transitions rather than merely changing a common relaxation scale. We examine how that redistribution affects work, heat, and repeated-cycle fluctuations under locally rate-matched and equal-output comparisons.

We investigate this mechanism using an anharmonic collective Otto engine with a longitudinal $J_z^2$ interaction. The interaction separates neighboring Dicke transition frequencies while preserving their collective matrix elements and a common energy eigenbasis throughout the ideal cycle. The isolated work strokes are consequently frictionless, whereas each thermal contact becomes a multi-edge transport process. Local detailed balance fixes the ratio of upward and downward rates at each transition frequency, while the reservoir spectrum determines their relative kinetic strengths~\cite{Davies1974,BreuerPetruccione2002,AlickiGelbwaserKlimovsky2015}. For the frequency-only work protocol, a given net excitation displacement produces the same work regardless of its location on the ladder, whereas the corresponding hot heat depends on the transitions involved. Reservoir shaping can therefore redistribute thermodynamically distinct transport currents without changing the Gibbs fixed points of the thermal contacts. We refer to this mechanism as \emph{spectral routing}.

The signed currents across the Dicke edges determine mean heat and work, whereas the gross activity counts the transitions driven by each contact, including crossings that cancel in the net current. This distinction makes the redistribution of kinetic activity measurable independently of its thermodynamic consequences. Because incomplete thermal contacts also retain information across cycle boundaries, changing the transition pathways can alter fluctuations accumulated over repeated cycles.

Superconducting circuits provide the ingredients for frequency-selective dissipation, including tunable electromagnetic environments, collective interactions, and passive microwave filters~\cite{Jones2013TunableEnvironment,Wang2019CavityAttenuator,Yan2023BroadbandPurcell,Blais2021,Song2019,Wang2020Superradiance}. Microwave thermometry and engineered thermal contacts provide complementary experimental capabilities~\cite{Scigliuzzo2020MicrowaveThermometry,Uusnakki2026,Aamir2025}. We analyze six transmons coupled to a common microwave bus and connected to equilibrium thermal reservoirs by passive filters. Microscopic dressed-spectrum calculations determine the transition energies and matrix elements, while symmetry-resolved population dynamics describe finite-time transport. At equal mean work and fixed integrated filter coupling, the structured configuration delivers approximately $24\%$ more output per thermal-contact time than the fastest sampled flat reference, while maintaining comparable long-time squared precision.

\noindent\emph{Contributions and scope.} We establish spectral routing as a kinetic control of finite-time thermodynamic transport in an anharmonic collective Otto engine without changing the Gibbs fixed points of its thermal contacts. Exact edge-current identities, a quantitative gross-activity diagnostic, locally rate-matched controls, and the noninteracting null limit identify the edge-selective mechanism. Repeated-cycle statistics and a memory-erasure control show how routing changes long-time work precision through correlations between cycles, while equal-output comparisons quantify its resource advantage. We delineate the mesoscopic range in which collective enhancement and spectral resolution remain compatible, and examine refrigeration and resilience of one-cycle precision to local-noise-induced spin-sector mixing. The circuit-QED implementation uses microscopic dressed spectra and symmetry-reduced population dynamics with a fixed passive-filter coupling budget; the full-space local-noise calculations are a separate robustness test.

\begin{figure*}[t!]
\centering
\includegraphics[width=0.98\textwidth]{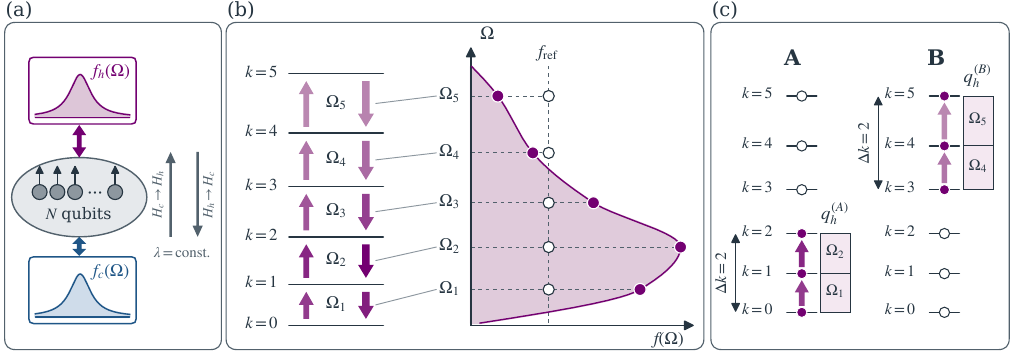}
\caption{\label{fig:mechanism}
Spectral routing in an anharmonic collective Otto engine. 
(a) Collective-spin working medium coupled to frequency-dependent hot and cold thermal reservoirs with spectral responses \(f_h(\Omega)\) and \(f_c(\Omega)\). The vertical arrows denote the isolated work strokes \(H_c \to H_h\) and \(H_h \to H_c\) at fixed interaction strength \(\lambda\). 
(b) The anharmonic interaction separates neighboring Dicke transition frequencies \(\Omega_k\). A structured reservoir response (solid curve) can therefore weight different ladder edges independently, in contrast to a flat reference (dashed line). 
(c) Two hot-contact transfer pathways with the same net excitation displacement \(\Delta k = 2\). Route A traverses edges 1 and 2, while route B traverses edges 4 and 5. For the fixed-\(\lambda\) protocol, both yield the same extracted work \(w = 2(\omega_h - \omega_c)\) but absorb different amounts of hot heat, \(q_h^{(A)} = \Omega_{h,1} + \Omega_{h,2}\) and \(q_h^{(B)} = \Omega_{h,4} + \Omega_{h,5}\). Frequency-selective dissipation changes the statistical weights of these thermodynamically distinct transfers without changing the thermal fixed points.}
\end{figure*}

\section{Collective Otto cycle and stochastic routing}
\label{sec:model-routing}

We consider a finite collective-spin working medium with $N$ identical two-level constituents, whose adjacent transitions are collectively enhanced and spectrally distinguishable. In the ideal collective model, the dynamics is restricted to the permutation-symmetric sector $j=N/2$. The Dicke states $\lvert k\rangle$, $k=0,\ldots,N$, satisfy
\begin{equation}
J_z\lvert k\rangle
=\left(k-\frac{N}{2}\right)\lvert k\rangle .
\end{equation}
Here $J_z=\tfrac12\sum_{n=1}^{N}\sigma_z^{(n)}$ measures the collective population imbalance, and $J_\pm=\sum_{n=1}^{N}\sigma_\pm^{(n)}$ raise or lower the collective excitation number by one. The adjacent transition $k\to k-1$ is denoted edge $k$. Its collective matrix element is
\begin{equation}
A_k=\left|\langle k-1\lvert J_-\rvert k\rangle\right|^2
=k(N-k+1),
\label{eq:collective-factor}
\end{equation}
so that $A_k$ depends only on $k$ and $N$, not on the transition frequency~\cite{Dicke1954,Jaseem2023,GarcezEtAlCollective2026}.

For a finite device, we consider the Hamiltonian family
\begin{equation}
H(\omega,\lambda)=\omega J_z+\lambda J_z^2,
\label{eq:finite-hamiltonian}
\end{equation}
where $\omega$ denotes the bare transition frequency of the two-level constituents and $\lambda$ controls the collective nonlinear interaction responsible for the anharmonic spectrum. At the cold and hot endpoints, denoted by $r\in\{c,h\}$, the level energies and adjacent transition frequencies are
\begin{align}
E_{r,k}&=\omega_r m_k+\lambda_r m_k^2,
\qquad m_k=k-\frac{N}{2},\nonumber\\
\Omega_{r,k}&=E_{r,k}-E_{r,k-1}
=\omega_r+\lambda_r(2k-N-1).
\label{eq:finite-spectrum}
\end{align}
Here, $\Omega_{r,k}$ denotes the Bohr frequency associated with the transition between Dicke levels $k$ and $k-1$, in units where $\hbar=1$.

The nonlinear interaction makes the transition frequencies depend on the excitation number, with neighboring frequencies separated by $\Omega_{r,k+1}-\Omega_{r,k}=2\lambda_r$. Thus, $\lambda_r$ controls the spectral separation of the Dicke transitions without changing their collective matrix elements in Eq.~\eqref{eq:collective-factor}. The positivity of all adjacent gaps, together with the special case in which all gaps are rescaled by the same factor between the cold and hot endpoints, is examined in Appendix~\ref{app:spectral-geometry}. The Hamiltonian in Eq.~\eqref{eq:finite-hamiltonian} contains no implicit system-size normalization; two distinct prescriptions for varying $N$ are introduced in Sec.~\ref{sec:mesoscopic-controls}.

The cycle consists of two isolated work strokes separated by hot and cold thermal contacts~\cite{Quan2005,Quan2007,RezekKosloff2006}. During an ideal work stroke,
\begin{equation}
H(t)=\omega(t)J_z+\lambda(t)J_z^2,
\end{equation}
so that $[H(t),H(t')]=0$ for all $t,t'$. The Dicke populations are therefore preserved exactly by the effective work-stroke dynamics. Finite laboratory stroke durations enter through the physical cycle time
\begin{equation}
\tau_{\rm cyc}=\tau_{u,c\to h}+\tau_h+\tau_{u,h\to c}+\tau_c.
\label{eq:cycle-time}
\end{equation}
Rates normalized by the thermal-contact time are referred to as contact-time rates.

The mean output power is $\mathcal P=W/\tau_{\rm cyc}$, where $W$ is the mean extracted work per cycle. We distinguish this physical cycle power from the
contact-time-normalized work output $W/(\tau_h+\tau_c)$, which isolates the time spent in thermal contact with the reservoirs. The two coincide only when the durations of the isolated work strokes are negligible.

During contact $r$, a structured passive thermal reservoir induces transitions between neighboring Dicke levels. In the frequency-resolved weak-coupling description, the downward and upward rates connecting levels $k$ and $k-1$ are
\begin{align}
R_{r,k}^{\downarrow}
&=\gamma_r A_k f_r(\Omega_{r,k})
\bigl[n_r(\Omega_{r,k})+1\bigr],\nonumber\\
R_{r,k}^{\uparrow}
&=\gamma_r A_k f_r(\Omega_{r,k})
n_r(\Omega_{r,k}),
\label{eq:thermal-rates}
\end{align}
where $n_r(\Omega)=[\exp(\beta_r\Omega)-1]^{-1}$ is the Bose occupation and $f_r(\Omega)>0$ is the dimensionless reservoir response. These rates are obtained from the diagonal restriction of a Davies weak-coupling thermal generator of Gorini--Kossakowski--Lindblad--Sudarshan (GKLS) form~\cite{Davies1974,GKS1976,Lindblad1976,BreuerPetruccione2002}. They satisfy the local detailed-balance relation,
\begin{equation}
\frac{R_{r,k}^{\uparrow}}{R_{r,k}^{\downarrow}}
=e^{-\beta_r\Omega_{r,k}},
\label{eq:kms}
\end{equation}
so that spectral shaping changes the speed of individual transitions while preserving their thermal equilibrium ratios. Consequently, the Gibbs distribution remains stationary under each thermal contact.

For Dicke-diagonal states, the population $p_k$ changes through transitions to and from neighboring levels,
\begin{align}
\dot p_k={}&
R_{r,k+1}^{\downarrow}p_{k+1}
+R_{r,k}^{\uparrow}p_{k-1}-\bigl(R_{r,k}^{\downarrow}
+R_{r,k+1}^{\uparrow}\bigr)p_k,
\label{eq:population-dynamics}
\end{align}
with transitions outside the Dicke ladder omitted at the boundaries. The first two terms represent population entering level $k$, while the last two represent population leaving it. These equations define the tridiagonal rate matrix $L_r$ through $\dot{\mathbf p}=L_r\mathbf p$, where $\mathbf p=(p_0,\ldots,p_N)^T$. The population-transfer matrix for a thermal contact of duration $\tau_r$ is then $K_r(\tau_r)=e^{L_r\tau_r}$.

The frequency-resolved description requires neighboring transition frequencies to remain separated relative to dissipative broadening, while the Markovian approximation requires sufficiently short bath correlation times. Near-degenerate transitions may instead require grouped or partially secular generators
~\cite{SchallerBrandes2008,Trushechkin2021}.

The mechanism is particularly transparent for the frequency-only work protocol, in which $\omega$ changes between the cold and hot endpoints while $\lambda_h=\lambda_c\equiv\lambda$. If the working medium enters the hot contact in Dicke level $i$ and leaves it in level $j$, the extracted work and absorbed hot heat for this trajectory are
\begin{align}
w(i,j)&=\Delta\omega\,(j-i),\nonumber\\
q_h(i,j)&=(j-i)
\bigl[\omega_h-N\lambda+\lambda(i+j)\bigr],
\label{eq:fixed-lambda-work-heat}
\end{align}
where $\Delta\omega=\omega_h-\omega_c$ and the extracted work is taken as positive. For a given net excitation displacement $j-i$, the work is independent of the initial level, whereas the hot heat depends on which part of the anharmonic ladder is traversed. Figure~\ref{fig:mechanism} illustrates this distinction.

Let $X_k$ denote the signed number of crossings of edge $k$ during the hot contact, counted positively for upward transitions. Its mean $\Phi_k=\langle X_k\rangle$, evaluated in the stationary Otto cycle, is
\begin{equation}
\Phi_k=\int_0^{\tau_h}\!dt\,
\left[
R_{h,k}^{\uparrow}p_{k-1}(t)
-R_{h,k}^{\downarrow}p_k(t)
\right].
\label{eq:edge-current}
\end{equation}
The two terms are the upward and downward probability currents across edge $k$. In the stationary cycle, the cold contact reverses the mean population change produced during the hot contact, so its net crossing of each edge is $-\Phi_k$. The mean energy exchanges therefore satisfy
\begin{align}
Q_h&=\sum_{k=1}^{N}\Omega_{h,k}\Phi_k,\nonumber\\
W&=\sum_{k=1}^{N}
(\Omega_{h,k}-\Omega_{c,k})\Phi_k,\nonumber\\
Q_c&=-\sum_{k=1}^{N}\Omega_{c,k}\Phi_k.
\label{eq:edge-energy-balances}
\end{align}
For fixed $\lambda$, the frequency difference $\Omega_{h,k}-\Omega_{c,k}=\Delta\omega$ is the same on every edge. Consequently, $W=\Delta\omega\sum_k\Phi_k$, while $Q_h$ depends on the distribution of the edge currents through their individual frequencies $\Omega_{h,k}$. By changing where population moves along the ladder, reservoir shaping can alter the heat absorbed from the hot bath even when the net excitation transfer, and hence the extracted work, remains unchanged.

To separate spectral routing from a uniform change in relaxation speed, we compare the structured reservoir with two reference contacts. The first has a flat spectral response, $f_r(\Omega)=1$, so that its transition rates retain the collective factors and thermal occupations but have no additional frequency-dependent weighting.

The second reference contact is a flat reservoir whose response is matched to the structured reservoir at a reference transition, chosen here as edge $k=1$:
$f_r^{\rm loc}(\Omega)=f_r(\Omega_{r,1})$ for all $\Omega$.
All transition rates are then multiplied by the same constant. Let $K_r^{\rm loc}$ and $K_r^{\rm flat}$ denote the population-transfer matrices of the locally matched reservoir and the unit-response flat reservoir, respectively. Introducing the dimensionless contact exposure $u_r=\gamma_r\tau_r$, we obtain
\begin{equation}
K_r^{\rm loc}(u_r)
=K_r^{\rm flat}
\!\left[f_r(\Omega_{r,1})u_r\right].
\label{eq:retiming-identity}
\end{equation}
A uniform rescaling of the reservoir response is equivalent to changing the contact duration. Spectral routing becomes possible when the reservoir changes the relative transition rates along the Dicke ladder, redistributing population among its edges.

In the noninteracting limit $\lambda=0$, all adjacent transitions have the same frequency. The structured reservoir then multiplies every transition rate by the same factor, and its population-transfer probabilities coincide with those of the locally matched flat reservoir. At matched effective contact exposures, this equivalence extends to the stationary trajectory distribution, all finite-cycle work--heat statistics, and repeated-cycle cumulants, not only the mean work and heat. The corresponding generator identities are given in Appendix~\ref{app:routing-controls}.

Figure~\ref{fig:mechanism} illustrates the connection between the cycle, the resolved spectrum, and the stochastic consequences. The work strokes change the energies assigned to occupied Dicke states without moving population between them; the contacts move population, and each adjacent transition samples the reservoir at its own Bohr frequency. For the frequency-only protocol, two transfers with the same $\Delta k$ therefore contribute the same work but can carry different hot heat because they traverse different parts of the anharmonic ladder. Reservoir structure changes the probabilities of these thermodynamically inequivalent routes while Eq.~\eqref{eq:kms} remains satisfied on every edge. This separation between an unchanged contact equilibrium and altered finite-time routing is tested quantitatively in Sec.~\ref{sec:frontier-memory}.

For the stochastic thermodynamic description, we place the cycle boundary immediately before the cold-to-hot work stroke. After repeated cycles, the working medium reaches a periodic steady state. Its population distribution at the chosen boundary, $\mathbf p_{\mathrm{ss}}=(p_{\mathrm{ss},0},\ldots,p_{\mathrm{ss},N})^T$, satisfies
\begin{equation}
\mathbf p_{\mathrm{ss}}
=K_cK_h\mathbf p_{\mathrm{ss}},
\qquad
\mathbf 1^T\mathbf p_{\mathrm{ss}}=1,
\label{eq:stationary-cycle}
\end{equation}
where $\mathbf 1$ denotes the vector with all entries equal to one. The distribution $\mathbf p_{\mathrm{ss}}$ is stationary at the cycle boundary, even though the populations evolve during each thermal contact.

An individual cycle is specified by the sequence $i\to j\to\ell$, where $i$ denotes the Dicke level before the cold-to-hot work stroke, $j$ denotes the level after the hot contact, and $\ell$ denotes the level after the cold contact. The probability of this sequence is
\begin{equation}
\mathbb P(i,j,\ell)
=p_{\mathrm{ss},i}(K_h)_{ji}(K_c)_{\ell j}.
\label{eq:path-probability}
\end{equation}

To characterize work and heat fluctuations, we use the two-point-measurement (TPM) approach, in which energy measurements at the process endpoints determine the stochastic energy change~\cite{TalknerLutzHanggi2007,EspositoHarbolaMukamel2009,CampisiHanggiTalkner2011}. Since the work-stroke Hamiltonians commute, the isolated strokes preserve the occupied Dicke level. The levels $i$ and $j$ therefore determine the extracted work $w_{ij}$ and the hot heat $q_{h,ij}$ of an individual cycle through the energy differences defined above. The corresponding one-cycle joint distribution is
\begin{equation}
P_1(w,q_h)
=\sum_{i,j}p_{\mathrm{ss},i}(K_h)_{ji}
\delta(w-w_{ij})
\delta(q_h-q_{h,ij}).
\label{eq:joint-distribution}
\end{equation}
Here, $w$ and $q_h$ denote possible values of the stochastic work and hot heat, while the delta functions assign probabilities to the discrete outcomes allowed by the Dicke spectrum. The marginal work distribution,
\begin{equation}
P_1(w)=\int dq_h\,P_1(w,q_h),
\end{equation}
gives the probability of each work outcome independently of the heat absorbed. The final level $\ell$ does not enter this distribution because the cold contact contributes neither to the extracted work nor to the hot heat, but it determines the state entering the next cycle. When the cold contact does not fully thermalize the working medium, memory of previous energy exchanges can persist, producing correlations between the work and heat fluctuations of successive cycles even after the engine reaches its periodic steady state.

For $M$ consecutive cycles, we define the accumulated work and hot heat as $W_M=\sum_{n=1}^{M}w_n$ and $Q_{h,M}=\sum_{n=1}^{M}q_{h,n}$, respectively, where $w_n$ and $q_{h,n}$ denote the stochastic work and hot heat exchanged in cycle $n$. Their joint statistics are encoded in the characteristic function $\chi_M(\xi_W,\xi_Q)=\langle e^{i(\xi_W W_M+\xi_Q Q_{h,M})}\rangle$, where $\xi_W$ and $\xi_Q$ are auxiliary counting parameters associated with work and hot heat. Derivatives with respect to these parameters yield the moments and correlations of the accumulated energy exchanges.

Starting from the stationary-cycle distribution $\mathbf p_{\mathrm{ss}}$, the characteristic function is given by $\chi_M(\xi_W,\xi_Q)=\mathbf 1^T\mathcal T(\xi_W,\xi_Q)^M \mathbf p_{\mathrm{ss}}$, where the tilted one-cycle propagator is
\begin{equation}
\mathcal T(\xi_W,\xi_Q)
=K_cD(\xi_W,\xi_Q)K_hD(-\xi_W,-\xi_Q).
\label{eq:tilted-map}
\end{equation}
Here, $D(\xi_W,\xi_Q)=\exp\left( i[\xi_W G+\xi_Q E_h] \right)$, with $G=\mathrm{diag}(E_{h,k}-E_{c,k})$ and $E_h=\mathrm{diag}(E_{h,k})$. The diagonal matrix $G$ contains the energy differences between the hot and cold Hamiltonians, while $E_h$ contains the hot-endpoint energies. The two counting factors assign the phase $\text{exp}\left( i\left[ \xi_W w_{ij}+\xi_Q q_{h,ij} \right] \right)$ to each hot-contact transition $i\to j$, thereby recording its contributions to the stochastic work and hot heat.

For a large number of cycles, the eigenvalue of $\mathcal T$ that approaches unity at zero counting parameters gives the long-time cumulants through derivatives of its logarithm~\cite{EspositoHarbolaMukamel2009,MohantaAgarwalla2023}. The asymptotic work variance per cycle is
\begin{equation}
\begin{aligned}
D_W^{(\infty)}
&=\lim_{M\to\infty}
\frac{\operatorname{Var}(W_M)}{M}\\
&=\operatorname{Var}(w_1)
+2\sum_{s=1}^{\infty}
\operatorname{Cov}(w_1,w_{1+s}),
\end{aligned}
\label{eq:long-time-diffusion}
\end{equation}
where $\operatorname{Cov}(w_1,w_{1+s})$ denotes the correlation between work outcomes separated by $s$ cycles. The first term is the variance of a single stationary cycle, while the sum contains the correlations between different cycles. These correlations can increase or reduce the long-time work fluctuations~\cite{Watanabe2017,GarcezEtAlCollective2026}.

We characterize the reliability of work extraction by the squared precision, defined as the squared mean work divided by its variance. Writing $W=\langle w\rangle$ for the mean extracted work per cycle, we distinguish the single-cycle and long-time precisions,
\begin{equation}
\Pi_W^{(1)}=\frac{W^2}{\operatorname{Var}(w_1)},
\qquad
\Pi_W^{(\infty)}=\frac{W^2}{D_W^{(\infty)}}.
\label{eq:squared-precision}
\end{equation}
The first quantity measures fluctuations within an individual stationary cycle, while the second includes correlations accumulated across repeated cycles. Time-normalized precision is written as $\Pi_W^{(\infty)}/\tau$, with the corresponding duration $\tau$ specified explicitly.

We compare structured and flat reservoirs through the dimensionless gain
\begin{equation}
\mathcal G_X=\frac{X_{\rm str}}{X_{\rm flat}},
\label{eq:gain-definition}
\end{equation}
where $X_{\rm str}$ and $X_{\rm flat}$ are the values of the same observable or performance metric obtained with the structured and flat contacts, respectively. For a nonzero reference value, $\mathcal G_X=1$ indicates equality, and $\mathcal G_X>1$ an enhancement of $X$ by the structured reservoir. Such an enhancement is favorable when larger values of $X$ are desirable, as for extracted work or precision.

Appendix~\ref{app:cycle-statistics} collects the exact finite-cycle and counting-field constructions. These quantities are used in Sec.~\ref{sec:frontier-memory} to examine how the finite-time work output and its statistical precision change with the reservoir response and contact duration.

\section{Routing the mean--fluctuation frontier}
\label{sec:frontier-memory}

\begin{figure*}[t!]
 \centering
 \includegraphics[width=1\textwidth]{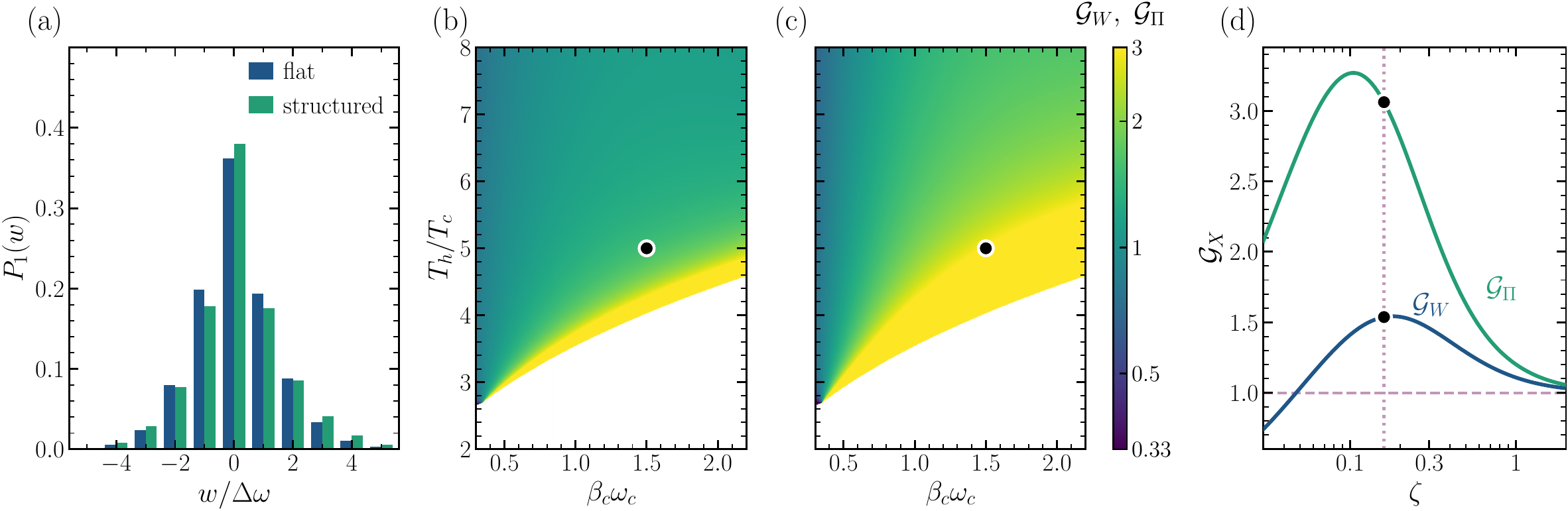}
 \caption{\label{fig:structured-faq}
Comparison of work statistics and precision gains for structured versus flat reservoirs.
Structured and flat contacts are compared at the same total reduced exposure
$U=0.8/N$ and hot fraction $u_h/U=0.45$, with the working medium of
Eq.~\eqref{eq:fig2-model-point}. The structured response is a band-normalized
Lorentzian centered at $0.80\,\omega_r$, with dimensionless width $\zeta=0.16$
in (a)--(c). In (b)--(d), the black dot marks the operating point
$\beta_c\omega_c=1.5$, $T_h/T_c=5$, $\zeta=0.16$.
(a)~Stationary one-cycle work distribution at the operating point for the flat
(blue) and structured (green) contacts; structuring changes the probabilities
of the discrete work outcomes, not their allowed values.
(b)~Mean-work gain $\mathcal G_W=W_{\rm str}/W_{\rm flat}$ over the thermal plane.
(c)~Long-time squared-precision gain
$\mathcal G_\Pi=\Pi^{(\infty)}_{W,\rm str}/\Pi^{(\infty)}_{W,\rm flat}$ over the
same plane. Panels (b) and (c) share one logarithmic color scale with
$\mathcal G=1$ at its center; gains outside the displayed range saturate at the
end colors. White regions mark thermal parameters where the two protocols do not
both have positive stationary mean work.
(d)~$\mathcal G_W$ and $\mathcal G_\Pi$ at the marked thermal point versus the
Lorentzian width $\zeta$. The dashed line marks $\mathcal G=1$, the flat
reference, and the dotted line marks $\zeta=0.16$, the width used in (a)--(c).}
\end{figure*}

We compare structured and flat thermal contacts at the same dimensionless contact exposures $u_r=\gamma_r\tau_r$, keeping the allocation between the hot and cold contacts fixed. This comparison identifies changes in work output and long-time precision arising from the frequency dependence of the reservoir response. In Fig.~\ref{fig:structured-faq}, $\mathcal G_W$ denotes the ratio of stationary mean extracted work, while $\mathcal G_{\Pi}$ denotes the ratio of long-time squared work precision, as defined in Eq.~\eqref{eq:gain-definition}. Values above unity favor the structured contact, whereas values below unity favor the flat reference.

The representative finite-size calculations use
\begin{equation}
N=12,\qquad \omega_c=1,\qquad \omega_h=2,\qquad \lambda=0.08,
\label{eq:fig2-model-point}
\end{equation}
with a total dimensionless contact exposure $U=u_h+u_c=0.8/N$ and hot-contact fraction $u_h/U=0.45$ for both protocols.

The structured reservoir has a Lorentzian spectral response $f_r(\Omega)$, peaked at the frequency $\Omega=0.80\,\omega_r$. This frequency dependence favors Dicke transitions near the spectral peak, while transitions farther away experience weaker coupling to the reservoir. Here $\kappa$ denotes the half-width at half-maximum of the Lorentzian peak and determines the range of transition frequencies strongly coupled to the reservoir. We define the dimensionless spectral width
\begin{equation}
\zeta=\frac{\kappa}{\omega_r},
\label{eq:zeta-width}
\end{equation}
so that a smaller $\zeta$ corresponds to a narrower reservoir response and a greater spectral selectivity. Panels (a)--(c) use $\zeta=0.16$; the spectral normalization and the complete set of numerical parameters are given in Appendix~\ref{app:results-provenance}.

Panel (a) shows the stochastic content of the comparison. The commuting work strokes and the allowed Dicke-state displacement fix the support of $P_1(w)$, while reservoir structure reweights the probabilities on that support. Spectral routing therefore redistributes trajectory probability across a fixed work spectrum, so that the mean, width, and probability of rare outcomes can respond differently. Related interacting Otto engines show that coupling-induced changes in level structure can increase or decrease work fluctuations~\cite{Anka2024}. For fixed $\lambda$, the level energies are unchanged between reservoir protocols; spectral shaping instead changes the probabilities assigned to the same allowed work outcomes. We count both upward and downward crossings of each Dicke edge during the stationary hot contact. Writing $\mathcal C_k$ for the mean total number of crossings of edge $k$, its normalized gross activity is $q_k=\mathcal C_k/\sum_{\ell=1}^{N}\mathcal C_\ell$; unlike the signed net current $\Phi_k$, it also counts transitions that cancel in net transport. Relative to the locally rate-matched flat contact, the routing distance is
\begin{equation}
\mathcal R_{\rm edge}=\frac{1}{2}\sum_{k=1}^{N}
\left|q_k^{\rm str}-q_k^{\rm loc}\right|.
\label{eq:main-routing-distance}
\end{equation}
At the representative operating point, $\mathcal R_{\rm edge}\simeq0.184$: edges $k=3$--$5$ account for $61\%$ of the structured contact's gross transition activity, compared with $43\%$ for the flat contact matched at edge $k=1$ on each thermal contact. Thus the reservoir changes the distribution of edge activity rather than merely rescaling all rates. The distance measures kinetic redistribution, not net heat current or a thermodynamic monotone; its definition and exact $\lambda=0$ null limit are detailed in Appendix~\ref{app:routing-controls}.

Panels (b) and (c) separate two questions that a single optimized operating point cannot answer. The mean-work map shows where the same contact resource produces more useful output, whereas the precision map shows where repeated operation produces a larger mean-squared signal per asymptotic variance. The common axes show that these are related but distinct objectives: the thermal parameters that favor output do not determine the magnitude of the precision gain. White regions lie outside the comparison in which both protocols operate as engines because at least one protocol has nonpositive stationary work.

At fixed contact exposures and identical reservoir coupling scales, the structured and flat protocols have the same thermal-contact durations. For identical work-stroke durations, the mean-work gain $\mathcal G_W$ also represents a gain in cycle-averaged output power. Spectral routing can increase the work produced per unit time without requiring longer thermal contacts.

Panel (d) tests whether the result depends on one specially chosen spectral width. In the broad-response limit the structured reservoir becomes effectively flat and both gains approach unity. Narrowing the response first increases both gains, but the dependence is nonmonotonic: making the bath arbitrarily narrow does not continue to improve the engine, and sufficiently strong selectivity can reduce the mean-work gain even while the long-time precision remains enhanced. The useful regime is an intermediate one in which the reservoir resolves and reweights several thermally active Dicke edges rather than either averaging over them or selecting too small a part of the ladder.

We introduce two equal-output controls in addition to the fixed-resource comparison in Fig.~\ref{fig:structured-faq} to test whether a flat contact can recover the same mean output by waiting longer. The first keeps the hot (cold) allocation fixed and retimes the flat protocol until its mean work matches the structured value. The second allows both flat contact durations to vary while imposing the same mean-work constraint. Appendix~\ref{app:results-provenance} collects their numerical values.

With the same hot (cold) allocation, routing improves both the one-cycle and long-time squared precision at matched mean work, and the larger change occurs in the repeated-cycle quantity. Over the broader retimed flat family, a sufficiently long flat contact can achieve a larger one-cycle squared precision than the structured protocol. Within the equal-output family considered here, the structured protocol retains the larger long-time squared precision, and also the larger long-time squared precision per unit contact exposure. At the representative $N=12$ point, its $\Pi_W^{(\infty)}/U$ exceeds the largest value over the examined equal-work flat family by a factor of $2.68$ (Appendix~\ref{app:results-provenance}). This factor concerns reduced thermal-contact exposure $U$, not the full physical cycle duration. These comparisons show that routing shifts the finite-time output--precision frontier: routing can improve repeated-cycle precision at matched output even when a longer flat protocol has the larger one-cycle precision.

The different one-cycle and long-time precision rankings arise from correlations between successive cycles. The dimensionless memory factor
\begin{equation}
\mathcal M_W=
\frac{D_W^{(\infty)}}{\operatorname{Var}(w_1)},
\qquad
\Pi_W^{(\infty)}
=\frac{\Pi_W^{(1)}}{\mathcal M_W},
\label{eq:work-memory-factor}
\end{equation}
quantifies their contribution. A value $\mathcal M_W<1$ indicates that the accumulated intercycle covariance reduces the asymptotic variance. At the representative point, $\mathcal M_W=0.450$ for the structured reservoir and $0.665$ for the equal-exposure flat contact. Both protocols exhibit negative intercycle correlations, but spectral shaping strengthens their contribution: despite having a larger single-cycle work variance, the structured protocol has a smaller asymptotic variance. When the flat contact is retimed to reproduce the same mean work at fixed hot (cold) allocation, its memory factor becomes $0.716$. The structured protocol's long-time precision advantage of $1.82$ combines a one-cycle precision gain of $1.14$ with a memory-factor gain of $1.59$. A marginal-preserving randomization at the cycle boundary removes the intercycle covariance tail while leaving the stationary one-cycle statistics unchanged. Independently, extending the cold contact drives $\mathcal M_W$ toward unity as the boundary state is more completely reset (Appendix~\ref{app:results-provenance}). These controls isolate cycle-to-cycle memory from the one-cycle work distribution. Negative intercycle correlations have previously been identified as a mechanism for stabilizing microscopic cyclic heat engines~\cite{Xu2022}. The reservoir spectrum provides an additional control: changing the edge-dependent thermal rates modifies the covariance correction without changing the work-stroke Hamiltonians.

Spectral routing also affects the heat-to-work conversion efficiency, $\eta=W/Q_h$. For the frequency-only protocol, the work per net transferred excitation is fixed by $\Delta\omega$, whereas the absorbed hot heat depends on the transitions carrying that transfer. Introducing the effective hot transition energy
\begin{equation}
\overline{\Omega}_h
=\frac{\sum_k\Omega_{h,k}\Phi_k}{\sum_k\Phi_k},
\qquad
\eta=\frac{\Delta\omega}{\overline{\Omega}_h},
\label{eq:routing-efficiency}
\end{equation}
shows that redistributing the edge currents changes the energetic cost of work extraction. At the representative operating point, the structured reservoir increases the efficiency from $0.559$ to $0.587$ at equal contact exposure. When the flat reservoir is retimed to reproduce the same mean work, its efficiency reaches $0.576$. Spectral routing can improve efficiency by reducing the hot heat required for a given output, although this improvement is not guaranteed for every reservoir spectrum.

The fixed-resource map and the equal-output controls address complementary questions. Figure~\ref{fig:structured-faq} shows where and how strongly reservoir structure changes the stochastic engine under the same contact budget. After the flat reference is retimed to the same mean output, spectral routing still shifts the repeated-cycle output--precision frontier, isolating the edge-selective contribution from an overall contact-rate change.

\section{Mesoscopic window and control landscape}
\label{sec:mesoscopic-controls}

\begin{figure}[t!]
\centering
\includegraphics[width=0.45\textwidth]{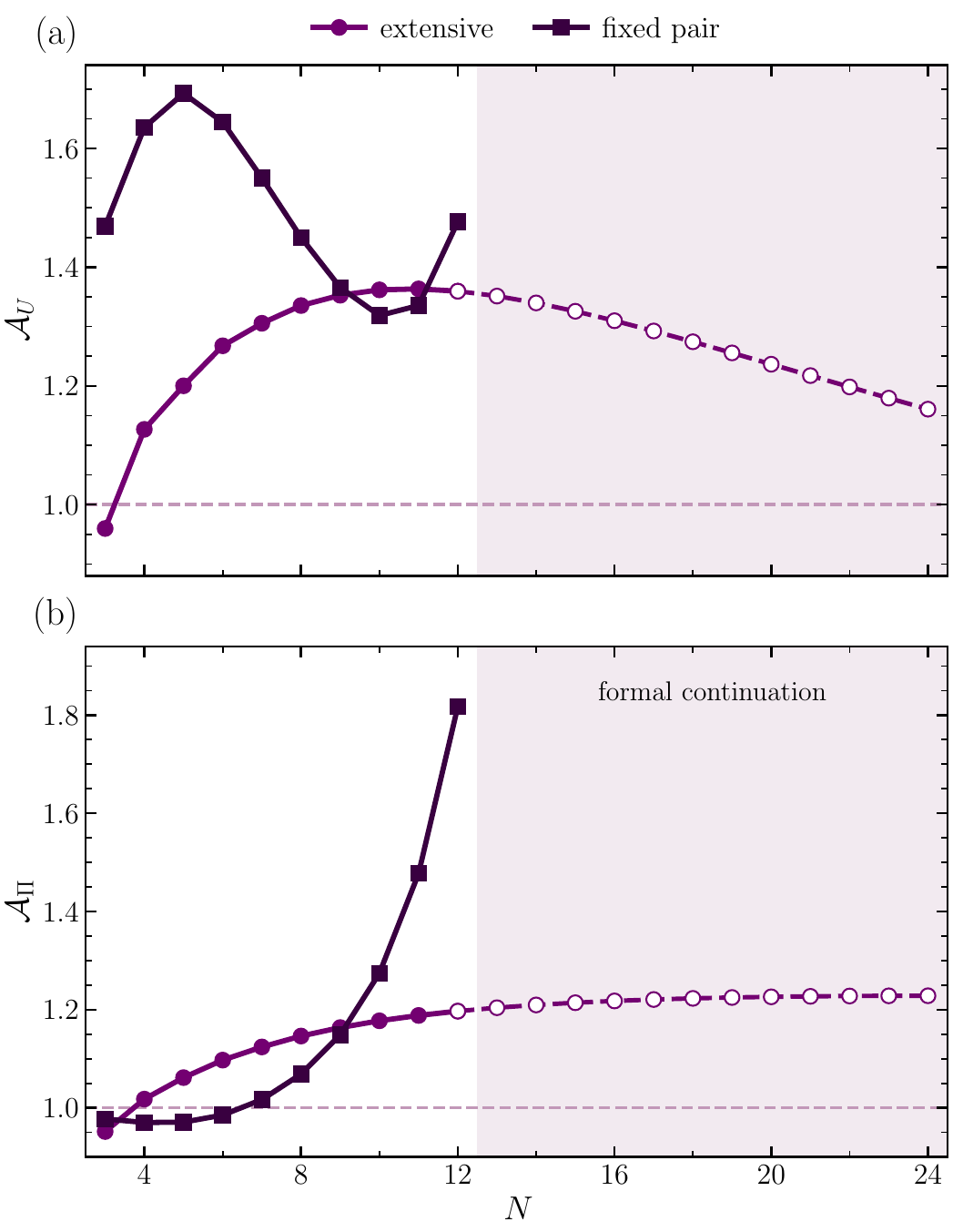}
\caption{\label{fig:mesoscopic-window}
Finite-size dependence of the structured-reservoir advantage at equal work.
Both panels compare the structured protocol with a flat reservoir retimed to the
same mean work, using $\omega_c=1$, $\omega_h=2$, $\beta_c=1.5$, $\beta_h=0.3$,
total reduced exposure $U=0.8/N$, and hot-contact fraction $0.45$. The structured
contacts use the same band-normalized Lorentzian as in
Fig.~\ref{fig:structured-faq}, with center $0.80\,\omega_r$ and width
$\zeta=0.16$. Circles denote the extensive continuation
$\lambda_N=(11/15)/N$, and squares denote the fixed-pair continuation
$\lambda=0.08$.
(a)~Equal-work exposure factor $\mathcal A_U=U_{\rm flat}/U_{\rm str}$.
(b)~Equal-work long-time squared-precision factor $\mathcal A_\Pi$.
In both panels, values above unity (dashed horizontal line) indicate an
advantage of the structured contact for the corresponding quantity. For the
extensive branch, the filled symbols and solid curve ($N\le12$) satisfy the
decade-scale resolved-channel criterion at the reference microscopic coupling.
Open symbols, dashed curves, and the shaded region ($N>12$) show the formal
continuation beyond this fixed-$\gamma$ window; retaining the same
resolved-Davies interpretation there requires weaker coupling and longer
physical contact times. The fixed-pair branch is shown through $N=12$, near the
upper end of the positive-gap range where the lowest cold transition softens
(Appendix~\ref{app:mesoscopic-boundary-support}).}
\end{figure}

The results above show that spectral shaping can improve work extraction and precision in a finite collective engine. Whether this advantage persists as the working medium size increases depends on how the collective interaction changes with system size. Increasing $N$ introduces more Dicke levels and strengthens the collective transition matrix elements, while the anharmonic spectrum determines which transitions remain accessible to the thermal reservoirs. These competing effects can make the routing advantage nonmonotonic with $N$.

We retain the Hamiltonian in Eq.~\eqref{eq:finite-hamiltonian} but allow its nonlinear interaction strength $\lambda$ to vary with the number of constituents. Denoting this size-dependent coupling by $\lambda_N$, we consider two distinct prescriptions.

In the first prescription, the interaction strength decreases inversely with system size,
\begin{equation}
\lambda_N=\frac{11}{15N},
\label{eq:kac-family}
\end{equation}
so that the collective interaction energy scales extensively with $N$. We refer to this as the extensive continuation. The separation between neighboring transition frequencies, $2\lambda_N$, decreases as $1/N$.

In the second prescription, the interaction strength is held fixed as the number of constituents increases,
\begin{equation}
\lambda_N=0.08,
\label{eq:fixed-pair-family}
\end{equation}
which corresponds to a fixed pairwise interaction. In this fixed-pair continuation, the frequency separation remains constant, but the lowest cold transition,
\begin{equation}
\Omega_{c,1}=1-0.08(N-1),
\label{eq:fixed-pair-lowest-gap}
\end{equation}
decreases with $N$ and eventually limits the accessible system sizes through the positive-gap condition.

We adjust the total contact exposure of the flat reservoir to reproduce the work obtained with the structured reservoir, keeping the hot (cold) exposure allocation fixed. We denote the corresponding total dimensionless exposures by $U_{\rm flat}$ and $U_{\rm str}$, respectively, and define
\begin{equation}
\mathcal A_U=\frac{U_{\rm flat}}{U_{\rm str}},
\qquad
\mathcal A_\Pi=
\frac{\Pi^{(\infty)}_{W,\rm str}}
{\Pi^{(\infty)}_{W,\rm flat}},
\label{eq:equal-work-size-advantages}
\end{equation}
where both precision values are evaluated at the same mean extracted work. Hence $\mathcal A_U>1$ means that the structured reservoir achieves the same output with less total contact exposure, while $\mathcal A_\Pi>1$ indicates higher long-time squared work precision under this equal-work comparison.

Figure~\ref{fig:mesoscopic-window} shows that the equal-work exposure advantage extends across a broad finite-size range in both interaction continuations. Along the extensive branch, $\mathcal A_U$ rises from near unity at the smallest sizes, reaches its largest values in the intermediate regime, and then decreases in the formal large-$N$ continuation. The long-time precision factor behaves differently, increasing gradually from below unity at the smallest size to a positive gain through the resolved mesoscopic window.

The fixed-pair branch makes the mechanism crossover more visible. Its exposure advantage is already substantial at small $N$, whereas its long-time precision is initially slightly worse than the equal-work flat reference. The precision ratio crosses unity near the middle of the branch and then rises rapidly toward $N=12$. The $N=12$ point lies near a crossover where the finite-cycle and intercycle contributions become simultaneously favorable.

The two branches also have different physical boundaries. For $\lambda_N\propto1/N$, the interaction-induced edge spacing shrinks with size, so at fixed microscopic coupling the resolved-channel hierarchy is eventually lost even though the reduced map can still be evaluated. The dashed extensive branch in Fig.~\ref{fig:mesoscopic-window} continues the reduced finite-$N$ map beyond the fixed-coupling resolved window; extending the same channel-resolved interpretation there requires weaker coupling and longer physical contact times. For fixed $\lambda$, the limitation instead comes from the softening of the lowest cold transition. In the occupied low-excitation sector of the extensive continuation, the collective linewidth scales as $\Gamma_{\rm occ}\sim\gamma_N N$, while the adjacent-edge separation scales as $N^{-1}$. Maintaining a size-independent resolved-channel margin therefore requires $\gamma_N$ to scale as $N^{-2}$ or faster toward zero, with thermal-contact durations growing at least as $N$ to retain nontrivial dynamics. For fixed work per cycle, physical power is consequently at most of order $N^{-1}$ along this resolution-preserving sequence; this is not a bound on grouped-transition or partially resolved regimes. Appendix~\ref{app:mesoscopic-boundary-support} gives the derivation, validity ratios, and asymptotic checks.

Hamiltonian and reservoir control also change in relative importance with system size. In the frequency-only protocol, the work strokes change $\omega$ while keeping the interaction strength fixed, so every Dicke transition frequency changes by the same amount $\Delta\omega$. Varying the interaction strength between the cold and hot endpoints introduces an additional control over the energy exchanged during each stroke. Interaction modulation is an established resource for improving Otto-engine performance~\cite{Boubakour2023}. We distinguish this control over the energetic weights of transitions from the kinetic reweighting provided by the reservoir spectra. For a population transfer $i\to j$, the extracted work becomes
\begin{equation}
w_{ij}=(j-i)
\bigl[\Delta\omega+\Delta\lambda(i+j-N)\bigr],
\label{eq:general-work-control}
\end{equation}
where $\Delta\lambda=\lambda_h-\lambda_c$.
The additional term makes the work depend on which Dicke levels participate in the transfer. Modulating the interaction can change the work associated with different parts of the ladder, whereas varying $\omega$ alone shifts every transition frequency by the same amount.

The thermal contacts offer another control. Instead of changing the energy associated with a transition, reservoir shaping changes its probability. The hot and cold reservoirs may share the same normalized spectral profile or have their spectral centers and widths adjusted independently. We compare these possibilities with the integrated spectral response of each contact held fixed, so that differences arise from the distribution of the available spectral weight. The work-stroke comparison also constrains the energy changes of the thermally active transitions. Appendix~\ref{app:mesoscopic-boundary-support} gives the corresponding constraints and numerical results.

Table~\ref{tab:control-hierarchy-ratios} shows how the relative benefits of Hamiltonian and reservoir control change with size. In the fixed-pair family at $N=4$, independently shaping the hot and cold spectra increases the maximum mean work by $8\%$, while jointly varying $\omega$ and $\lambda$ has almost no effect. At $N=12$ the ordering reverses: Hamiltonian modulation increases maximum work by $22\%$, compared with only $0.5\%$ for independent spectral shaping; the extensive family shows the same shift. Combining the controls brings no further work gain at this fixed-pair point, but raises the long-time squared-precision gain from about $72\%$ to $83\%$. Work and precision need not improve together: for the extensive family at $N=12$, independent spectral shaping increases maximum work by $3.7\%$ while reducing long-time squared precision by about $12\%$. These gains are relative to the frequency-only, shared-spectrum baseline with equal per-contact spectral budgets. Each strategy is optimized for mean work, and its precision is evaluated at that maximum-work point, not optimized independently. Hamiltonian modulation changes the energetic weights of transfers, while reservoir shaping controls the relative transition activity that realizes them.

In the stationary cycle, the net population transferred across each Dicke edge during the hot contact is canceled by the corresponding transfer during the cold contact. The two reservoirs can nevertheless produce different amounts of upward and downward transition activity while maintaining this balance. Adjusting their spectral responses independently changes how the required population transfer is realized in each contact. We refer to this possibility as \emph{contact-specific spectral specialization}.

The finite-size results show how the usefulness of these controls depends on the accessible Dicke spectrum. Reservoir shaping is particularly effective when several thermally occupied transitions can be addressed separately, while interaction modulation changes the energy assigned to transfers across the ladder. We examine the same reservoir-routing mechanism in refrigeration and in a noisy superconducting device.

\section{Thermodynamic transfer and passive circuit implementation}
\label{sec:consequences-device}

The same edge-selective kinetics can also improve refrigeration. When the cycle is operated to extract heat from the cold reservoir, spectral shaping changes which transitions carry the cooling current and how their fluctuations accumulate across cycles. Collective refrigeration is established, including a superradiant many-qubit absorption refrigerator~\cite{Kloc2021}. We test whether the same finite-time spectral-routing mechanism is useful when the prescribed output is cooling rather than work. For the $N=8$ example, the structured reservoir achieves the same mean cooling at $U=0.100$ as a retimed flat reference at $U\simeq0.190$, requiring about $1.90$ times less contact exposure. Its long-time cooling squared precision is about $14\%$ higher, while the probability of an unfavorable cold-contact heat transfer falls from $0.226$ to $0.205$. The corresponding coefficient of performance, $Q_c/|W|$, is approximately $0.440$ for the structured contact and $0.448$ for the equal-cooling flat reference: the kinetic and precision advantages do not imply a gain in every thermodynamic metric. The operating parameters and equal-cooling comparison are collected in Appendix~\ref{app:refrigerator-transfer}. The engine and refrigerator exploit the same frequency-selective population transport, although the resulting thermodynamic advantages depend on the operating mode.

\begin{figure*}[t!]
\centering
\includegraphics[width=0.91\textwidth]{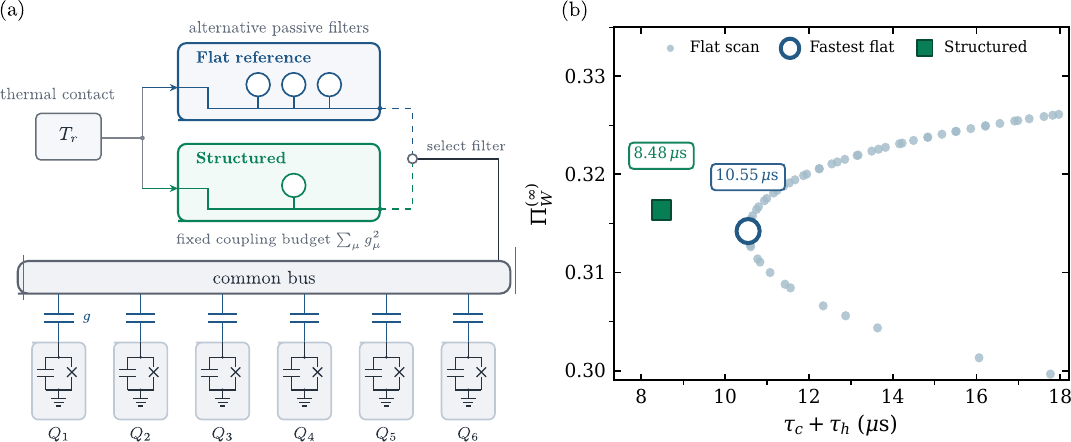}
\caption{\label{fig:passive-device}
Finite-time performance of a passive reservoir-shaping device at matched mean work. 
(a) Circuit schematic showing six transmons coupled to a shared microwave bus with coupling \(g = 40\,\mathrm{MHz}\). The thermal contacts are implemented via passive filters, comparing an approximately flat three-pole reference against a structured single-pole reservoir. Both filter designs operate under an identical integrated coupling budget, \(\sum_\mu g_{r\mu}^2 = (0.160\,\mathrm{MHz})^2\). 
(b) Long-time squared work precision as a function of total thermal-contact time, evaluated at temperatures \(T_c = 0.10\,\mathrm{K}\) and \(T_h = 0.50\,\mathrm{K}\), with local coherence times \(T_1 = T_2 = 30\,\mu\mathrm{s}\). The pale data points trace the equal-work solutions for the flat reference, with the open circle indicating the fastest sampled flat configuration and the filled square marking the structured operating point. At a matched mean work of \(W/h \simeq 7.03\,\mathrm{MHz}\), the structured device achieves the target precision in just \(8.48\,\mu\mathrm{s}\)—significantly faster than the \(10.55\,\mu\mathrm{s}\) required by the best flat reference. The resulting long-time squared precisions are nearly identical (\(0.3164\) for the structured device vs. \(0.3142\) for the flat one), demonstrating that spectral routing shortens the required thermal-contact time without sacrificing stability.}
\end{figure*}

We test whether the engine-side advantage survives local relaxation and dephasing that populate lower-total-spin sectors. For $N=6$, we evolve the stationary cycle in the full $2^6$-dimensional spin Hilbert space and compare structured and flat contacts at identical durations and local-noise parameters. Local relaxation and dephasing produce substantial spin-sector mixing: immediately before the hot contact, the structured cycle has a median lower-spin population of about $31\%$, reaching $70\%$ across the sampled domain. The one-cycle squared-precision gain survives in approximately $93\%$ of the $478$ cases in which more than half of the population lies outside the symmetric $j=3$ sector, and in approximately $96\%$ of the $1673$ comparisons in which the structured contact has greater lower-spin occupation than the flat reference. Its absolute one-cycle work variance is larger in all $2586$ accepted cases: wherever precision increases, the larger mean work more than compensates for the additional variance. The advantage therefore persists without preferential confinement to the symmetric Dicke manifold. The sampled parameter domain, inclusion criteria, and spin-sector diagnostics are given in Appendix~\ref{app:fullspace-noise}.

Collective heat-current enhancement in circuit QED has been studied in the presence of a parasitic local reservoir~\cite{Andolina2024}. We ask whether passive spectral shaping improves a finite-time Otto cycle at matched work and a fixed integrated filter-coupling budget.

We consider six superconducting transmons coupled capacitively to a common microwave bus, as shown in Fig.~\ref{fig:passive-device}(a). The bus mediates the effective interaction between the transmons and provides a channel through which the working medium couples to external microwave reservoirs. A microscopic transmon--bus calculation gives the dressed transition frequencies and matrix elements used in the thermal-contact model. Within the populated spectral region, these energies agree with the effective collective-spin description while retaining the sector-dependent shifts of the microscopic circuit. Appendix~\ref{app:transmon-bus} gives the Hamiltonian parameters and the dressed-spectrum comparison.

The thermal reservoirs consist of equilibrium microwave loads coupled through passive damped filter modes. Each mode contributes a Lorentzian spectral response whose center frequency, linewidth, and coupling strength determine which dressed transitions are most strongly driven. We compare an approximately flat three-pole reference with a structured single-pole reservoir. Both configurations have the same integrated filter-coupling strength on each thermal contact. For the eliminated-filter Lorentzians, $\int_{-\infty}^{\infty}J_r(\nu)\,d\nu=4\pi^2\sum_\mu g_{r\mu}^{2}$, so the fixed sum of squared filter couplings fixes the integrated spectral weight, not the rate of each individual dressed transition. Spectral shaping redistributes this resource among the accessible transition frequencies without requiring a coherent sideband pump. The corresponding spectral densities and filter parameters are listed in Appendix~\ref{app:passive-ports}.

Figure~\ref{fig:passive-device}(b) shows the finite-time benefit of passive spectral shaping. At equal mean work, the structured reservoir reduces the total thermal-contact time from $10.55\,\mu\mathrm{s}$ to $8.48\,\mu\mathrm{s}$, corresponding to a $24\%$ increase in contact-time-normalized work output. The gain in physical cycle power depends on the durations of the isolated work strokes and any additional switching overhead. The implementation establishes a finite-time output advantage without assuming instantaneous laboratory control. The long-time squared work precision changes by less than $1\%$; combining this near-equality with the shorter contact time gives an approximately $25\%$ advantage in $\Pi_W^{(\infty)}/(\tau_h+\tau_c)$ for the structured device. This is long-time precision per thermal-contact time, not precision per complete laboratory cycle. The flat-reference points correspond to different allocations of the hot and cold contact durations, each adjusted to reproduce the structured mean work. The advantage is reaching nearly the same output--precision operating point with shorter thermal contacts.

This benefit is specific to the finite-time comparison. At longer contact durations, some flat-reference solutions achieve higher intrinsic precision. The one-cycle squared precisions of the selected operating points are nearly identical, while the flat reference has a smaller negative-work probability. The different fluctuation measures need not rank the reservoirs in the same way. Under the fixed passive coupling budget, the principal operational advantage is the reduction in contact time at a prescribed mean work and comparable long-time precision.

The dressed transitions lie near $5.2\,\mathrm{GHz}$, whereas the work strokes produce much smaller changes in their frequencies. The routed operating point has a gross heat-to-work efficiency of approximately $0.24\%$. The low value reflects the large microwave carrier energy relative to the work extracted during the cycle; it does not include external switching or other system-level costs. The corresponding energy accounting is presented in Appendix~\ref{app:passive-comparison}.

The local-noise and passive-device calculations address complementary aspects of the implementation. The local-noise calculation evolves the full six-spin density matrix and shows a one-cycle precision advantage under mixing of total-spin sectors. In the passive-filter comparison, the microscopic transmon--bus spectrum supplies dressed transition energies and matrix elements to a 16-state $(j,m)$ population model. Local relaxation and dephasing in this reduced model are obtained by averaging over the multiplicities of the spin sectors. The passive-device result concerns finite-time performance within a dressed-population description, rather than a complete open-system simulation of the transmons, bus, and filter modes. The reduction and its frequency-resolution checks are given in Appendix~\ref{app:transmon-bus}.

The full-space calculation shows that the finite-time routing mechanism survives local noise and population transfer between total-spin sectors. The dressed-circuit calculation demonstrates the same mechanism using passive laboratory-frequency reservoirs and a fixed microscopic coupling resource. These results link the ideal collective-engine mechanism to more realistic dissipative dynamics and a physically specified circuit architecture.

\section{Discussion and conclusion}
\label{sec:discussion}

The influence of reservoir structure on the collective Otto cycle arises from the separation between transition energies and transition rates. The longitudinal interaction resolves neighboring Dicke transitions without changing their collective matrix elements, while local detailed balance preserves the Gibbs state of each thermal contact. A frequency-dependent reservoir can therefore redistribute finite-time transition activity without changing the equilibrium toward which the working medium relaxes. The collective matrix elements provide a strongly nonuniform kinetic landscape, although spectral routing itself is not restricted to collective systems. Gross edge activity characterizes the redistribution of transitions, whereas signed edge currents determine their contributions to work and heat. When the work strokes modulate only the bare transition frequency, each net excitation transfer contributes the same extracted work, but the associated hot heat depends on which edges carry that transfer. This distinction explains how reservoir shaping modifies heat-to-work conversion even when the working-medium Hamiltonian and the thermal fixed points remain unchanged.

The consequences extend beyond the work distribution of an individual cycle. At fixed thermal-contact exposure, the structured reservoir can increase both mean extracted work and long-time squared precision. The equal-output comparisons show that this improvement is not equivalent to uniformly accelerating relaxation. One-cycle and long-time precision need not rank the same protocols in the same order. Their difference is governed by the intercycle covariance, which depends on the population distribution retained across successive thermal contacts. Negative correlations between successive work outcomes are known to stabilize microscopic cyclic engines~\cite{Watanabe2017,Xu2022}; spectral routing provides a means of modifying these correlations through edge-dependent thermalization. Randomizing the state at each cycle boundary while preserving its stationary marginal removes intercycle correlations without changing the one-cycle statistics. Enhanced long-time precision can therefore coexist with a larger absolute work variance, so single-cycle and accumulated measures capture different aspects of operation.

The available range of spectral control depends on the collective spectrum. Increasing the number of constituents introduces additional transport pathways and strengthens collective transition amplitudes, but these benefits compete with the frequency resolution needed to distinguish neighboring edges. In the extensive interaction family, the transition spacing decreases with system size; in the fixed-pair family, the lowest cold transition progressively softens. The resulting mesoscopic window reflects two distinct spectral limitations rather than a universal optimal system size. Maintaining individually resolved transitions as their spacing contracts requires weaker coupling and longer physical contacts. Hamiltonian and reservoir controls also act differently within this window: varying $(\omega,\lambda)$ changes the work assigned to population transfers, whereas shaping the hot and cold spectra changes their relative kinetic weights. The relative advantage of these controls consequently depends on both system size and the performance objective.

The same transport mechanism operates beyond the ideal collective engine. In refrigeration, structured contacts reach the prescribed cooling output with less thermal-contact exposure and higher long-time cooling precision than the retimed flat reference at the operating point studied. Local relaxation and dephasing produce substantial spin-sector mixing. As shown in Fig.~\ref{fig:fullspace-noise-appendix}, the one-cycle squared-precision advantage persists
across much of the sampled parameter domain, including cases in which most of the population lies outside the symmetric $j=3$ sector. The structured reservoir also produces a larger absolute work variance throughout the accepted dataset, so the observed precision improvement is driven by increased mean work rather than suppressed fluctuations or preferential confinement to the symmetric Dicke manifold. This behavior is relevant to collective thermal machines in which exchange-symmetry sectors contribute differently to energy transport~\cite{Boeyens2025}.

The circuit-QED calculation places spectral routing under a microscopic coupling constraint. Six transmons interact through a common microwave bus, and their dressed transition energies and matrix elements determine the rates induced by passive thermal filters. The structured and approximately flat filters have the same integrated coupling budget. At equal mean work, the structured configuration requires $8.48\,\mu\mathrm{s}$ of thermal contact, compared with $10.55\,\mu\mathrm{s}$ for the fastest sampled flat configuration, while their intrinsic long-time squared precisions remain comparable. This corresponds to approximately $24\%$ greater work per unit thermal-contact time. The physical cycle power also depends on the durations of the isolated strokes and reservoir switching, whose inclusion in a fully dynamical circuit description would connect the present contact-time results to complete device operation.

The distribution of finite-time transition currents thus provides a thermodynamic control variable distinct from the overall relaxation rate. Anharmonicity makes the transitions spectrally accessible, collective coupling determines their kinetic strengths, and passive reservoir shaping controls how they participate in energy exchange. Through this redistribution, reservoir spectra influence heat conversion, fluctuations accumulated over repeated cycles, and the contact time required to deliver a prescribed output.

\section*{Data and code availability}
Data and numerical code supporting the findings of this study are available from the corresponding author upon reasonable request.

\section*{Acknowledgments}
This work was supported by the Coordenação de Aperfeiçoamento de Pessoal de Nível Superior - Brasil (CAPES) - Finance Code 001.

\subsection*{Artificial Intelligence Usage Declaration}
OpenAI ChatGPT (GPT-5.6 Sol) was used
for language editing, bibliographic organization, and code-assisted
analytical and numerical cross-checks. The authors specified the physical
model and validation criteria, checked the resulting calculations against
analytic identities and numerical convergence tests, and assume full
responsibility for the manuscript.

\appendix
\section{Spectral geometry and commuting work strokes}
\label{app:spectral-geometry}

For the finite-device Hamiltonian $H(\omega,\lambda)=\omega J_z+\lambda J_z^2$, the Dicke-state energy and adjacent gap are
\begin{equation}
E_k=\omega\left(k-\frac{N}{2}\right)+\lambda\left(k-\frac{N}{2}\right)^2,
\end{equation}
\begin{equation}
\Omega_k=\omega+\lambda(2k-N-1),\qquad k=1,\ldots,N.
\label{eq:appA-gap}
\end{equation}
Neighboring transitions are separated by $2\lambda$. The complete ladder is positive when
\begin{equation}
\omega>|\lambda|(N-1).
\label{eq:positive-gap-domain}
\end{equation}
This spectral condition is distinct from dynamical frequency resolution, which additionally compares the gap separations with dissipative linewidths and reservoir correlation times.

For two endpoints, define $s_k=\Omega_{h,k}/\Omega_{c,k}$. If $N\ge2$, a common positive scale $s_k=s$ for every edge is equivalent to
\begin{equation}
\omega_h=s\omega_c,\qquad \lambda_h=s\lambda_c.
\label{eq:homothetic-coefficients}
\end{equation}
On this homothetic line, $H_h=sH_c$ up to any physically irrelevant additive constant and every edge has the same Otto conversion ratio. For $N=1$, $J_z^2=I/4$ and the quadratic coefficient cannot be inferred from the single transition.

The edgewise deformation away from homothety follows from
\begin{equation}
s_{k+1}-s_k=
2\,\frac{\omega_c\lambda_h-\omega_h\lambda_c}
{\Omega_{c,k+1}\Omega_{c,k}}.
\end{equation}
Within the positive-gap domain, the sign of $\omega_c\lambda_h-\omega_h\lambda_c$ therefore fixes the orientation of the compression-ratio gradient across the ladder.

During an isolated stroke,
\begin{equation}
H(t)=\omega(t)J_z+\lambda(t)J_z^2,
\end{equation}
and $[H(t),H(t')]=0$ for arbitrary $t,t'$. The propagator is diagonal in the Dicke basis and preserves the Dicke populations exactly within the effective model. Finite laboratory stroke durations enter the cycle time, while leakage is treated explicitly in the full-space device calculation.

\section{Exact cycle thermodynamics and repeated-cycle statistics}
\label{app:cycle-statistics}

For contact $r\in\{c,h\}$, the population generator has off-diagonal entries
\begin{equation}
(L_r)_{k-1,k}=R_{r,k}^{\downarrow},\qquad (L_r)_{k,k-1}=R_{r,k}^{\uparrow},
\end{equation}
with the rates in Eq.~\eqref{eq:thermal-rates} and diagonal elements fixed by $\mathbf 1^TL_r=0$. The contact propagator is $K_r=e^{L_r\tau_r}$. The stationary cycle boundary is the normalized right eigenvector of
\begin{equation}
P=K_cK_h,
\qquad P\mathbf p_{\mathrm{ss}}=\mathbf p_{\mathrm{ss}}.
\end{equation}
Detailed balance on each isolated contact gives the Gibbs population of the corresponding endpoint Hamiltonian as its stationary state.

For a stationary path $i\to j\to\ell$,
\begin{equation}
\mathbb P(i,j,\ell)=p_{\mathrm{ss},i}(K_h)_{ji}(K_c)_{\ell j}.
\end{equation}
The endpoint energy changes are
\begin{align}
 w(i,j)&=[E_{h,j}-E_{c,j}]-[E_{h,i}-E_{c,i}],\nonumber\\
q_h(i,j)&=E_{h,j}-E_{h,i},\nonumber\\
q_c(j,\ell)&=E_{c,\ell}-E_{c,j}.
\end{align}
They obey the pathwise first law
\begin{equation}
E_{c,\ell}-E_{c,i}=q_h+q_c-w,
\end{equation}
and stationary averaging gives $W=Q_h+Q_c$.

For the quadratic endpoints, the general work increment is
\begin{equation}
 w(i,j)=(j-i)\bigl[\Delta\omega+\Delta\lambda(i+j-N)\bigr],
\end{equation}
while
\begin{equation}
q_h(i,j)=(j-i)\bigl[\omega_h-N\lambda_h+\lambda_h(i+j)\bigr].
\end{equation}
The frequency-only protocol has $\Delta\lambda=0$, which reduces the work to $\Delta\omega(j-i)$.

The one-cycle joint distribution is Eq.~\eqref{eq:joint-distribution}. For repeated cycles, define the diagonal matrices
\begin{equation}
G=\mathrm{diag}(E_{h,k}-E_{c,k}),\qquad E_h=\mathrm{diag}(E_{h,k})
\end{equation}
and
\begin{equation}
D(\xi_W,\xi_Q)=\exp\{i(\xi_WG+\xi_QE_h)\}.
\end{equation}
The tilted map is Eq.~\eqref{eq:tilted-map}, and for $M$ cycles
\begin{equation}
\mathcal Z_M(\xi_W,\xi_Q)=\mathbf 1^T\mathcal T(\xi_W,\xi_Q)^M\mathbf p_{\mathrm{ss}}.
\end{equation}
If $\zeta_0$ is the eigenvalue branch connected continuously to $\zeta_0(0,0)=1$, then
\begin{equation}
\psi(\xi_W,\xi_Q)=\lim_{M\to\infty}\frac{1}{M}\ln\mathcal Z_M=\ln\zeta_0.
\end{equation}
The first derivatives of $\psi$ give $W$ and $Q_h$; its Hessian gives the asymptotic work--heat covariance matrix. The work component can equivalently be written as
\begin{equation}
D_W^{(\infty)}=\operatorname{Var}(w_0)+2\sum_{q\ge1}\operatorname{Cov}(w_0,w_q).
\end{equation}

The squared precisions used in the manuscript are
\begin{equation}
\Pi_W^{(1)}=\frac{W^2}{\operatorname{Var}(w)},\qquad
\Pi_W^{(\infty)}=\frac{W^2}{D_W^{(\infty)}}.
\end{equation}
The negative-work probability is
\begin{equation}
P_-=\sum_i p_{\mathrm{ss},i}\sum_{j<i}(K_h)_{ji}
\end{equation}
when $\Delta\omega>0$ and all relevant hot gaps are positive. These quantities are kept separate because a protocol can improve one without improving the others.

For a complete-reset contact, $K_r=\boldsymbol\pi_r\mathbf 1^T$, successive cycles are independent and $D_W^{(\infty)}=\operatorname{Var}(w)$. This limit removes intercycle memory but does not remove the anharmonic spectral geometry.

\section{Routing controls and exact null limits}
\label{app:routing-controls}

The flat response is $f_r^{\rm flat}(\Omega)=1$. For a structured response, choose edge $k=1$ as a common reference and define
\begin{equation}
f_{r,1}=f_r(\Omega_{r,1}),\qquad f_r^{\rm loc}(\Omega)=f_{r,1}.
\end{equation}
Because the resolved thermal rates are linear in $f_r$,
\begin{equation}
L_r^{\rm loc}=f_{r,1}L_r^{\rm flat},
\end{equation}
which gives the exact retiming identity in Eq.~\eqref{eq:retiming-identity}. The structured generator can therefore be decomposed as
\begin{equation}
L_r^{\rm sh}=L_r^{\rm loc}+\Delta L_r^{\rm disp},
\end{equation}
where the edge-$k$ contribution to $\Delta L_r^{\rm disp}$ is proportional to
\begin{equation}
\delta f_{r,k}=f_r(\Omega_{r,k})-f_r(\Omega_{r,1}).
\end{equation}
The first term changes only the contact speed after retiming; the second distinguishes different edges and is the multi-edge routing contribution.

The mean number of crossings, counting both directions during a stationary hot contact, provides a direct measure of edge usage:
\begin{align}
\mathcal{C}_k&=\int_0^{\tau_h}\!dt\,
\bigl[R_{h,k}^{\uparrow}p_{k-1}(t)
     +R_{h,k}^{\downarrow}p_k(t)\bigr],\nonumber\\
q_k&=\frac{\mathcal{C}_k}{\sum_{\ell=1}^{N}\mathcal{C}_\ell}.
\label{eq:edge-activity}
\end{align}
Unlike the signed net crossing $\Phi_k$, the normalized activity $q_k$ is nonnegative even when some edges carry opposing net currents. The distance from the locally rate-matched flat reference is
\begin{equation}
\mathcal{R}_{\mathrm{edge}}=\frac{1}{2}\sum_{k=1}^{N}
\left|q_k^{\mathrm{str}}-q_k^{\mathrm{loc}}\right|.
\label{eq:edge-redistribution}
\end{equation}
Both distributions are calculated at the same hot and cold contact exposures, with the flat response on contact $r$ set to $f_r^{\mathrm{loc}}(\Omega)=f_r(\Omega_{r,1})$. A contact-wide rate rescaling is therefore included in the reference and is not mistaken for a frequency-selective effect. In particular, the structured and locally matched generators coincide for $\lambda=0$; hence $\mathcal{R}_{\mathrm{edge}}=0$ in that limit. At the representative Fig.~\ref{fig:structured-faq} point, $\mathcal{R}_{\mathrm{edge}}\simeq0.184$. This quantity measures a difference in normalized gross activity. It is not a thermodynamic monotone, nor does it measure the signed net heat current.

For the extensive continuation $\lambda_N=\chi/N$, a smooth positive spectrum, and fixed edge index $k$ as $N\to\infty$ (or a low-excitation range with bounded $k$),
\begin{equation}
\frac{f_r(\Omega_{r,k})}{f_r(\Omega_{r,1})}
=1+\frac{2\chi(k-1)}{N}
\left.\partial_\Omega\ln f_r(\Omega)\right|_{\Omega_{r,1}}+O(N^{-2}).
\end{equation}
The $O(N^{-2})$ remainder is not uniform for edges with $k=O(N)$. This relation diagnoses asymptotic spectral crowding within the stated low-excitation regime, while routing is defined by the edge-dependent rate reweighting that remains after local matching.

At $\lambda=0$, every edge on a given contact has frequency $\omega_r$. After the local exposure is matched, $L_r^{\rm sh}=L_r^{\rm loc}$ and the stationary population, trajectory law, work and heat distributions, finite-cycle generating functions, and repeated-cycle cumulants coincide exactly. This is the full-statistics noninteracting null control.

The other controls remove different physical ingredients. A fixed-total-exposure comparison redistributes contact time between the two isochores. A mean-work-matched flat protocol retimes the flat contacts until its useful mean output equals that of the structured protocol. A complete reset removes incoming-state memory. A homothetic deformation removes edge-dependent work-to-heat conversion ratios. None of these controls is interchangeable with the $\lambda=0$ spectral-addressability null.

\section{Refrigeration, local noise, and passive-device implementation}
\label{app:fullspace-device}

\subsection{Refrigerator transfer point}
\label{app:refrigerator-transfer}

For the refrigeration cycle, we use the same collective ladder with $N=8$, $\omega_c=1$, $\omega_h=2$, and $\lambda=\chi/N=11/120$, where $\chi=11/15$ specifies the extensive interaction scaling. The reservoir temperatures satisfy $\beta_c\omega_c=1.5$ and $T_h/T_c=2$. The total dimensionless contact exposure is $U=u_h+u_c=0.8/N=0.1$, with $u_h/U=0.45$ and $u_c/U=0.55$.

The structured response on contact $r$ is a Lorentzian centered at $c\omega_r$, with half-width $\zeta\omega_r$,
\begin{equation}
f_r(\Omega)\propto
\frac{(\zeta\omega_r)^2}
{(\Omega-c\omega_r)^2+(\zeta\omega_r)^2},
\qquad c=0.60,\quad \zeta=0.20.
\end{equation}
The proportionality constant gives unit mean response over $\Omega\in[\omega_r-|\chi|,\omega_r+|\chi|]$. The flat reference has $f_r(\Omega)=1$. Its total exposure is increased while keeping $u_h/U=0.45$ until the mean cooling matches that of the structured cycle.

We take $Q_c>0$ to denote heat extracted from the cold reservoir. At $U=0.1$, the structured contact extracts $Q_c\simeq0.0547$, with long-time cooling squared precision $\Pi_{Q_c}^{(\infty)}\simeq0.0202$ and cold-heating probability $P_{\mathrm{heat},c}\simeq0.205$. Here $P_{\mathrm{heat},c}$ denotes the probability that the cold contact transfers heat into, rather than out of, the cold reservoir. The equal-cooling flat reference requires $U\simeq0.190$, with $\Pi_{Q_c}^{(\infty)}\simeq0.0177$ and $P_{\mathrm{heat},c}\simeq0.226$. The structured reservoir therefore achieves the same cooling with approximately $47\%$ less contact exposure and $14\%$ higher long-time squared precision, while also reducing the probability of heating the cold reservoir.

\subsection{Full-space local noise and spin-sector populations}
\label{app:fullspace-noise}

Local relaxation and dephasing can transfer population between total-spin sectors. To study this effect, we evolve the complete $2^N$-dimensional spin density matrix with site-resolved noise channels and collective engineered thermal contacts. The population outside the permutation-symmetric sector is
\begin{equation}
 p_{\perp}=1-\operatorname{Tr}\!\left(P_{j=N/2}\rho\right),
\end{equation}
where $P_{j=N/2}$ projects onto the maximal-spin manifold. We evaluate this population in the stationary cycle immediately before the hot contact.

The numerical study uses $N=6$, with endpoint frequencies $\omega_c/(2\pi)=4.0\,\mathrm{GHz}$ and $\omega_h/(2\pi)=6.4\,\mathrm{GHz}$, $\lambda/\omega_c=0.12/N=0.02$, $\beta_c\omega_c=2$, and $\tau_c=\tau_h=8\,\mu\mathrm{s}$. The reservoir response $f_r(\Omega)$ is Lorentzian, centered at $\Omega=0.62\,\omega_r$ with half-width $0.22\,\omega_r$, and has the same integrated-band normalization as the flat reference. The scanned parameters are
\begin{align}
\gamma/2\pi
&\in\{1,1.5,2,2.5,3,4,5,6,8\}\,\mathrm{kHz},
\nonumber\\
T_h/T_c
&\in\{2,3,4,5,6,8,10,12\},
\nonumber\\
T_1/\mu\mathrm{s}
&\in\{50,75,100,150,200,300,500,750,1000\},
\nonumber\\
T_2/T_1
&\in\{0.5,1,1.5,2\}.
\end{align}

The frequency-resolved description is evaluated through three ratios comparing the neighboring transition spacing, reservoir linewidth, and smallest positive transition frequency with the largest noise-augmented broadening scale:
\begin{equation}
 \mathcal M_{\mathrm{sec}}=\frac{\Delta\Omega}{\Gamma_{\max}},\qquad
 \mathcal M_{\kappa}=\frac{\kappa}{\Gamma_{\max}},\qquad
 \mathcal M_{\mathrm{weak}}=\frac{\Omega_{\min}}{\Gamma_{\max}},
\end{equation}
where $\Delta\Omega=2|\lambda|$ is the adjacent-transition spacing within a collective ladder, $\kappa$ is the structured reservoir linewidth, $\Omega_{\min}$ is the smallest positive transition frequency used in the check, and $\Gamma_{\max}$ is the largest noise-augmented rate or broadening scale in the adopted generator. We require $\mathcal M_{\mathrm{sec}},\mathcal M_{\kappa}\ge10$ and $\mathcal M_{\mathrm{weak}}\ge20$ on each contact, together with a stationary-cycle residual not exceeding $2\times10^{-9}$ and positive-work operation for the compared cases. These criteria characterize frequency resolution within the collective ladders. Because local noise also couples different total-spin sectors, the within-ladder spacing is distinct from the separations between transitions belonging to different sectors.

Of the $2592$ parameter combinations, $2586$ meet all the recorded inclusion conditions. The six excluded combinations have $\gamma/2\pi=1\,\mathrm{kHz}$ and $T_h/T_c=2$, with
\[
\begin{aligned}
 (T_1/\mu\mathrm{s},T_2/T_1)\in\{&(200,0.5),(300,0.5),(300,1),\\
 &(500,0.5),(750,1.5),(1000,0.5)\}.
\end{aligned}
\]

\begin{figure*}[t]
\centering
\includegraphics[width=1\textwidth]{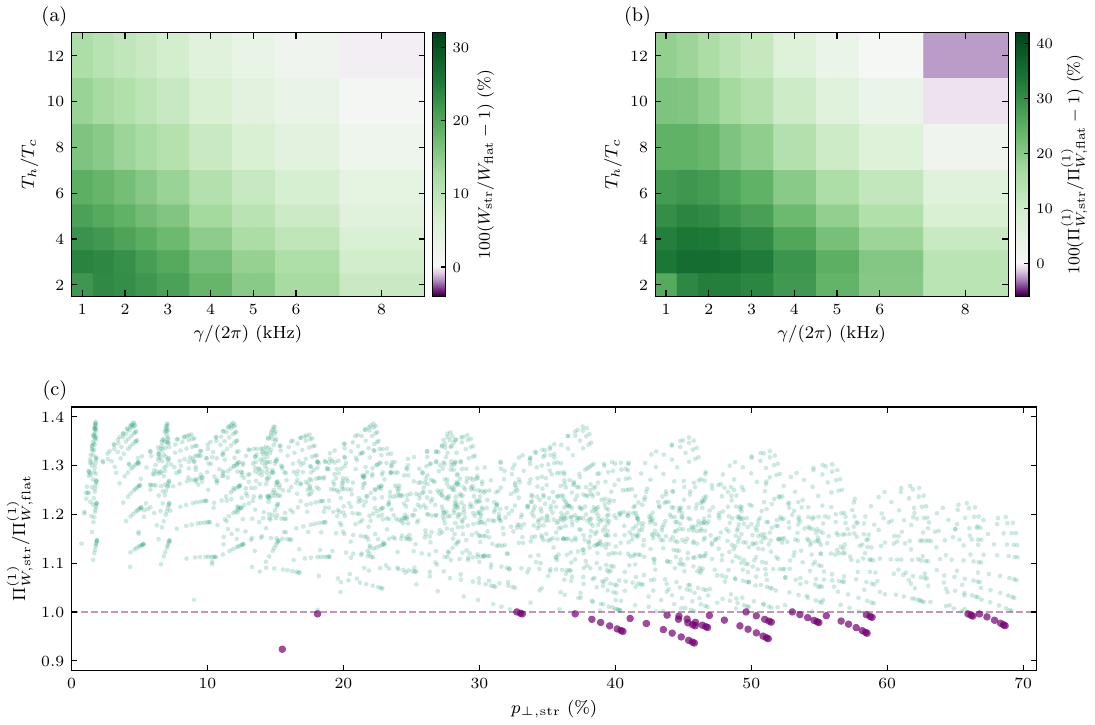}
\caption{\label{fig:fullspace-noise-appendix}
Full-space response to local relaxation and dephasing.
Panels (a) and (b) show the median percentage change in mean extracted work, $100(W_{\rm str}/W_{\rm flat}-1)$, and in one-cycle squared work precision, $100(\Pi^{(1)}_{W,\rm str}/\Pi^{(1)}_{W,\rm flat}-1)$, respectively, between the structured and flat contacts at equal durations; green indicates a gain of the structured contact and purple a loss. Medians are taken over the sampled $T_1$ and $T_2/T_1$ values at each collective-coupling and temperature-ratio point. Panel (c) compares the pre-hot lower-spin population of the structured contact, $p_{\perp,\rm str}$, with its one-cycle precision ratio relative to the flat reference; points above the dashed line ($\Pi^{(1)}_{W,\rm str}/\Pi^{(1)}_{W,\rm flat}=1$) are green and points at or below it are purple. The plotted data comprise the $2586$ accepted parameter combinations.}
\end{figure*}

For the structured contact, the median population outside $j=N/2$ is about $31\%$, reaching approximately $70\%$ across the sampled domain. Figure~\ref{fig:fullspace-noise-appendix}(c) shows that substantial occupation of lower-spin sectors does not eliminate the one-cycle precision advantage.
Among the $478$ cases with $p_{\perp,\mathrm{str}}>0.5$, approximately $93\%$ ($445$ cases) retain a precision gain relative to the flat reference. The gain also persists in approximately $96\%$ of the $1673$ comparisons where the structured contact populates lower-spin sectors more strongly than the flat reference. The improvement therefore does not require preferential occupation of the symmetric Dicke manifold.

The structured contact has a larger absolute one-cycle work variance throughout the accepted dataset. With $W=\langle w\rangle$ and $w$ the stochastic work outcome, the precision ratio satisfies
\begin{equation}
 \frac{\Pi_{W,\rm str}^{(1)}}{\Pi_{W,{\rm flat}}^{(1)}}
=\left(\frac{W_{\rm str}}{W_{\rm flat}}\right)^2
 \frac{\operatorname{Var}_{\rm flat}(w)}{\operatorname{Var}_{\rm str}(w)}.
\end{equation}
Whenever this ratio exceeds unity, the increase in mean extracted work more than compensates for the larger work variance. The full-space result is therefore a mean-output-driven one-cycle precision improvement within the tested noise domain.

\subsection{Microscopic transmon--bus spectrum}
\label{app:transmon-bus}

The implementation model contains six identical Duffing transmons coupled uniformly to a detuned bosonic bus. In cyclic-frequency units, its Hamiltonian is
\begin{align}
 \frac{H_{\mathrm{circ}}}{h}
 ={}&\sum_{i=1}^{6}\left[\nu_q a_i^\dagger a_i
 +\frac{K_q}{2}a_i^{\dagger2}a_i^2\right]\nonumber\\
 &+\nu_b b^\dagger b
 +g\sum_{i=1}^{6}(a_i^\dagger b+b^\dagger a_i),
\end{align}
where $a_i$ annihilates an excitation of transmon $i$, $b$ annihilates a bus photon, $K_q$ is the transmon anharmonicity, and $g$ is the bus--transmon coupling. The cold and hot work-stroke endpoints have $\nu_q=5.200$ and $5.216\,\mathrm{GHz}$, respectively, with $\nu_b=6.000\,\mathrm{GHz}$, $K_q=-220\,\mathrm{MHz}$, and $g=40\,\mathrm{MHz}$. The parameters of $H_{\mathrm{circ}}/h$ are expressed in cyclic-frequency units. A dressed transition
with energy difference $E_\alpha-E_\beta$ has cyclic frequency $\nu_{\alpha\beta}=(E_\alpha-E_\beta)/h$ and angular Bohr frequency $\Omega_{\alpha\beta}=2\pi\nu_{\alpha\beta}$.

Excitation-sector diagonalization gives an effective nonlinear interaction of approximately $11.65$--$11.88\,\mathrm{MHz}$. Over the selected dressed band, the maximum adjacent-gap residual relative to the effective collective-spin description is below $0.53\,\mathrm{MHz}$, and the weight outside the computational, bus-vacuum sector remains below about $3.3\%$. The uniform bus-mediated exchange also produces a $J^2$ contribution: it is constant in a fixed-$j$ manifold but changes the relative offsets of different spin sectors. Those offsets are retained when constructing the dressed transition energies.

The passive-filter model uses the permutation-invariant populations $p_{j,m}$, with one dressed energy for each $(j,m)$ level. For six spin-$1/2$ constituents, the sectors $j=0,1,2,3$ contain $7+5+3+1=16$ distinct population variables. Their multiplicities are $d_3=1$, $d_2=5$, $d_1=9$, and $d_0=5$, satisfying $\sum_j(2j+1)d_j=64$.

Let $P_{jm}$ project onto all $d_j$ copies of $(j,m)$. The local relaxation and dephasing rates follow from applying the sums of site-local dissipators to the normalized sector state $P_{jm}/d_j$ and projecting onto $P_{j'm'}$. The resulting population generator includes transfers between different total-spin sectors with their correct multiplicities. For permutation-symmetric local noise, this generator agrees with the corresponding population restriction of the full $64$-dimensional spin dynamics.

The dressed transition energies and bus-mediated matrix elements determine the passive thermal-contact rates, while the local dissipators retain their effective spin-space form.

The collective filter-coupling operators $J_\pm$ conserve total spin $j$ and connect neighboring $m$ levels within each sector. For permutation-invariant states diagonal in the $(j,m)$ populations, collective thermal contacts preserve this structure, while local relaxation and dephasing transfer population between sectors. The population dynamics is thus closed within the sector-resolved model.

The smallest separation between neighboring collective transitions within a sector is approximately $3.61\,\mathrm{MHz}$ for the cold endpoint and $3.68\,\mathrm{MHz}$ for the hot endpoint. Transitions belonging to different $j$ sectors can be separated by as little as $0.094\,\mathrm{MHz}$ and $0.097\,\mathrm{MHz}$, respectively. Since $J_\pm$ acts independently within each $j$ sector, these cross-sector near-degeneracies do not introduce interference between the collective transitions for the population states considered here. At the selected filter parameters, the minimum ratio of an intra-sector angular-frequency separation to the larger of the two summed thermal rates is approximately $11$.

\subsection{Passive thermal ports and coupling resource}
\label{app:passive-ports}

\begin{figure*}[t]
\centering
\includegraphics[width=0.94\textwidth]{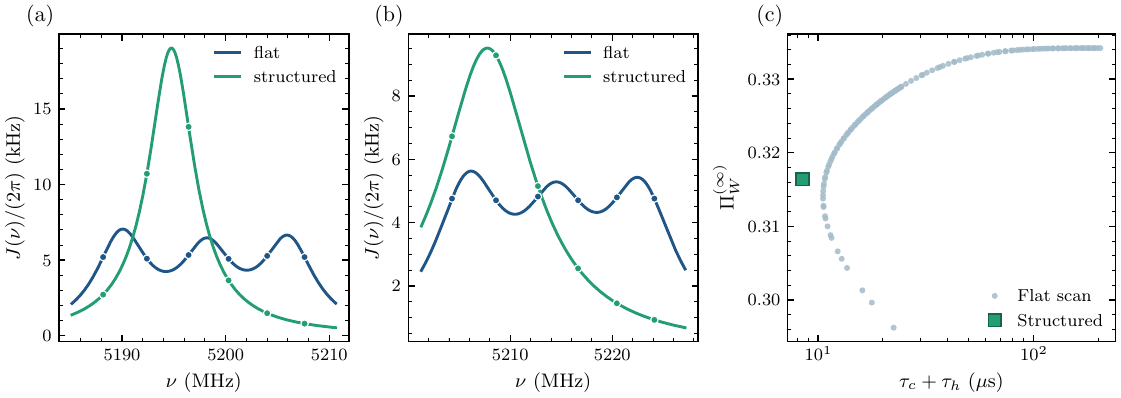}
\caption{\label{fig:passive-filter-details}
Passive microwave spectra and equal-work reference configurations.
Panels (a) and (b) show the cold- and hot-contact spectral densities $J(\nu)/2\pi$, respectively, of the approximately flat three-pole filter (blue) and the structured single-pole filter (green), under the same per-contact coupling budget $\sum_\mu g_\mu^2=(0.160\,{\rm MHz})^2$. Dots mark the spectral density at the six dressed transitions of the symmetric $j=3$ sector; the six additional transitions of the $j=1,2$ sectors also enter the population dynamics. Panel (c) shows the long-time squared work precision versus total thermal-contact time on a logarithmic scale, for the sampled flat-reservoir equal-work solutions (pale dots) and the structured operating point (green square), over the full contact-time range, complementing the focused comparison in Fig.~\ref{fig:passive-device}.}
\end{figure*}

The thermal contacts couple equilibrium microwave loads to the dressed transitions of the transmon system through damped passive filters. For filter modes $\mu$ with center frequency $\nu_{r\mu}$, full width at half maximum $\kappa_{r\mu}$, and coupling $g_{r\mu}$ to contact $r$, the eliminated-filter spectral density is
\begin{equation}
 J_r(\nu)=\sum_\mu 2\pi
 \frac{4g_{r\mu}^2\kappa_{r\mu}}
 {\kappa_{r\mu}^2+4(\nu-\nu_{r\mu})^2}.
 \label{eq:passive-filter-spectrum}
\end{equation}
The linewidth $\kappa_{r\mu}$ here is the \emph{full} width, unlike the half-width convention for the ideal Lorentzian responses used in the collective-ladder calculations. Integrating the model Lorentzians over the full frequency axis gives
\begin{equation}
 \int_{-\infty}^{\infty}J_r(\nu)\,d\nu
 =4\pi^2\sum_\mu g_{r\mu}^2.
\end{equation}
We compare the three-pole approximately flat reference with the one-pole structured reservoir at a fixed resource on each thermal contact,
\begin{equation}
 \sum_\mu g_{r\mu}^2=(0.160\,\mathrm{MHz})^2,
 \qquad r\in\{c,h\}.
 \label{eq:passive-coupling-budget}
\end{equation}
This constraint fixes the integrated spectral weight of each thermal contact while allowing its distribution across frequencies to vary. With the parameters in Table~\ref{tab:passive-filter-parameters}, the approximately flat three-pole reference distributes the coupling across the dressed-transition band, whereas the structured single-pole filter concentrates it near selected transitions [Fig.~\ref{fig:passive-filter-details}(a,b)]. The resulting differences in transition rates arise from spectral selectivity while the integrated coupling strength is held fixed.

\begin{table*}[t]
\centering
\caption{Passive-filter parameters for the equal-work device comparison. Center frequencies $\nu_{r\mu}$ and full linewidths $\kappa_{r\mu}$ are in MHz. The normalized weights are $w_{r\mu}=g_{r\mu}^2/\sum_\alpha g_{r\alpha}^2$.}
\label{tab:passive-filter-parameters}
\begin{ruledtabular}
\begin{tabular}{lccc}
Configuration & $\nu_{r\mu}$ & $\kappa_{r\mu}$ & $w_{r\mu}$\\
\hline
Flat, cold & $5189.96,\ 5198.19,\ 5206.05$ & $6,\ 6,\ 6$ & $0.3670,\ 0.2932,\ 0.3398$\\
Flat, hot & $5205.90,\ 5214.50,\ 5222.77$ & $8,\ 8,\ 8$ & $0.3695,\ 0.2812,\ 0.3493$\\
Structured, cold & $5194.77$ & $5.38$ & $1$\\
Structured, hot & $5207.76$ & $10.77$ & $1$\\
\end{tabular}
\end{ruledtabular}
\end{table*}

At $T_c=0.10\,\mathrm K$ and $T_h=0.50\,\mathrm K$, the equilibrium Bose occupations near $5.2\,\mathrm{GHz}$ are approximately $0.09$ and $1.55$, respectively. These thermal occupations determine the upward-to-downward rate ratios through detailed balance, while the spectral density $J_r(\nu)$ determines the coupling strength at each dressed transition.

\subsection{Equal-work comparison and energy accounting}
\label{app:passive-comparison}

The structured cycle uses $\tau_c=4.66\,\mu\mathrm{s}$ and $\tau_h=3.82\,\mu\mathrm{s}$ at $T_c=0.10\,\mathrm K$, $T_h=0.50\,\mathrm K$, and local $T_1=T_2=30\,\mu\mathrm s$. It produces $W_{\mathrm{str}}/h\simeq 7.03\,\mathrm{MHz}$, single-cycle squared work precision $\Pi_{W,\mathrm{str}}^{(1)}\simeq0.162$, and long-time squared precision $\Pi_{W,\mathrm{str}}^{(\infty)}\simeq0.316$. The flat reference at the same contact durations produces $W_{\mathrm{flat}}/h\simeq5.82\,\mathrm{MHz}$ and $\Pi_{W,\mathrm{flat}}^{(\infty)}\simeq0.263$.

For the equal-work comparison in Fig.~\ref{fig:passive-device}(b), the flat contact durations are varied independently at the same filter-coupling budget until $W_{\mathrm{flat}}=W_{\mathrm{str}}$. The shortest of the $144$ sampled flat solutions uses $\tau_c\simeq6.08\,\mu\mathrm{s}$ and $\tau_h\simeq4.47\,\mu\mathrm{s}$, totaling $10.55\,\mu\mathrm{s}$, compared with $8.48\,\mu\mathrm{s}$ for the structured operating point. At equal mean work, this corresponds to approximately $24\%$ greater work per unit thermal-contact time for the structured configuration. The long-time squared precisions are approximately $0.314$ and $0.316$, respectively, giving a time-normalized long-time precision ratio of about $1.25$ in favor of the structured cycle. Their single-cycle squared precisions are nearly equal. Across the broader equal-work family shown in Fig.~\ref{fig:passive-filter-details}(c), some flat solutions at longer contact times exceed the structured configuration's intrinsic long-time precision. The flat reference also has a smaller negative-work probability, $P_-=\Pr(w<0)$, at the representative equal-work operating point.

The laboratory-frame energy accounting includes heat exchanged through the passive thermal ports and local dissipative channels. The structured cycle absorbs approximately $Q_{h,\mathrm{th}}/h=2.90\,\mathrm{GHz}$ per cycle from the hot passive port. Together with $W/h\simeq7.03\,\mathrm{MHz}$, this gives a gross hot-port heat-to-work ratio $W/Q_{h,\mathrm{th}}\simeq 2.4\times10^{-3}$. The ratio refers to the passive-port thermal input; external switching and control costs enter the energy accounting of a complete device cycle.

For the structured single-pole filters, the largest effective coupling-to-linewidth ratio is $g_{\mathrm{eff}}/\kappa\simeq0.10$. The narrowest full linewidth, $\kappa=5.38\,\mathrm{MHz}$ in cyclic-frequency units, corresponds to a filter response time $1/(\pi\kappa)\simeq 0.06\,\mu\mathrm{s}$. This timescale is substantially shorter than the minimum thermal-contact duration of $3.82\,\mu\mathrm{s}$, placing filter relaxation and finite-time population transport on distinct timescales.

\section{Parameters, comparators, and technical diagnostics for the ideal-model results}
\label{app:results-provenance}

This appendix gives the numerical specification behind Fig.~\ref{fig:structured-faq} and the equal-output controls discussed in Sec.~\ref{sec:frontier-memory}. The thermal-plane comparison in Fig.~\ref{fig:structured-faq} varies the reservoir temperatures while keeping the Hamiltonian, contact exposure, and spectral parameters fixed as specified in Table~\ref{tab:fig2-parameters}. The population dynamics follow from Eq.~\eqref{eq:thermal-rates}, the stationary cycle from Eq.~\eqref{eq:stationary-cycle}, and the work statistics from Appendix~\ref{app:cycle-statistics}. Reduced contact exposures are used throughout this ideal-model comparison.

\begin{table}[t]
\caption{Fixed parameters for Fig.~\ref{fig:structured-faq}. The thermal variables $\beta_c\omega_c$ and $T_h/T_c$ are varied in panels (b,c); the starred point is the value shown in the last two rows.}
\label{tab:fig2-parameters}
\centering
\begin{tabular}{ll}
\hline\hline
Quantity & Value\\
\hline
$N$ & $12$\\
$\omega_c,\omega_h$ & $1,2$\\
$\lambda$ & $0.08$\\
total reduced exposure $U$ & $0.8/N$\\
hot fraction $u_h/U$ & $0.45$\\
Lorentzian center ratio & $0.80$\\
reference width $\zeta=\kappa/\omega_r$ & $0.16$\\
normalization half-band & $11/15$\\
star: $\beta_c\omega_c$ & $1.5$\\
star: $T_h/T_c$ & $5$\\
\hline\hline
\end{tabular}
\end{table}

The Lorentzian used in the figure is
\begin{equation}
f_r(\Omega)=C_r\frac{\kappa_r^2}{(\Omega-0.80\,\omega_r)^2+\kappa_r^2},
\qquad \kappa_r=\zeta\omega_r,
\label{eq:appE-lorentzian}
\end{equation}
with $C_r$ chosen separately for each contact so that the response has the same integrated band budget as the flat reference over the symmetric half-band $11/15$ about $\omega_r$. Panels (b,c) use the same structured spectrum while varying the thermal parameters. Only points for which both protocols have positive stationary mean work are displayed. Panel (d) holds $\beta_c\omega_c=1.5$ and $T_h/T_c=5$ fixed while varying $\zeta$.

At the starred structured point,
\begin{align}
\langle W\rangle_{\rm str}&\simeq0.110,\nonumber\\
\Pi_{W,\rm str}^{(1)}&\simeq5.36\times10^{-3},\nonumber\\
\Pi_{W,\rm str}^{(\infty)}&\simeq1.19\times10^{-2}.
\label{eq:appE-star-values}
\end{align}
These values use $U=0.8/N$ and $u_h/U=0.45$.

For the same-allocation equal-output comparator, the flat contact keeps $u_h/U=0.45$ and only its total exposure is changed until its mean work equals Eq.~\eqref{eq:appE-star-values}. The resulting precision ratios are
\begin{equation}
\frac{\Pi_{W,\rm str}^{(1)}}{\Pi_{W,\rm flat}^{(1)}}\simeq1.14,
\qquad
\frac{\Pi_{W,\rm str}^{(\infty)}}{\Pi_{W,\rm flat}^{(\infty)}}\simeq1.82.
\label{eq:appE-same-allocation}
\end{equation}
This fixed-allocation equal-output comparator isolates spectral shaping at unchanged hot (cold) exposure fraction.

The fully retimed flat family varies both contact durations while matching the same mean work. Over the equal-output family considered, the least flat exposure satisfies
\begin{equation}
\frac{U_{\rm flat,min}}{U_{\rm str}}\simeq1.47,
\qquad U_{\rm flat,min}\simeq0.0981,
\label{eq:appE-fast-flat}
\end{equation}
and the structured protocol obeys
\begin{equation}
\frac{\Pi_{W,\rm str}^{(1)}}{\max_{\rm flat}\Pi_W^{(1)}}\simeq0.77,
\qquad
\frac{\Pi_{W,\rm str}^{(\infty)}}{\max_{\rm flat}\Pi_W^{(\infty)}}\simeq1.70.
\label{eq:appE-strict-precision}
\end{equation}
Thus the strict family can exceed the structured one-cycle precision if substantially longer contacts are allowed, whereas the structured long-time precision remains larger over the equal-output family considered. Including reduced contact exposure gives
\begin{equation}
\frac{\Pi_{W,\rm str}^{(\infty)}/U_{\rm str}}
{\max_{\rm flat}[\Pi_W^{(\infty)}/U_{\rm flat}]}=2.68.
\label{eq:appE-precision-rate}
\end{equation}
These statements concern reduced contact exposure rather than the complete laboratory cycle time.

The repeated-cycle interpretation is supported by two independent renewal diagnostics. First, the structured protocol has a negative covariance tail, and the flat finite-contact reference is also anticorrelated; routing changes the magnitude of the memory correction rather than its sign class. Second, a marginal-preserving randomization of the end-of-cold-contact state removes the covariance tail while leaving all one-cycle statistics unchanged. Increasing the physical cold exposure provides the complementary dynamical check: $D_W^{(\infty)}/\operatorname{Var}(W)$ approaches unity as the boundary state is reset more completely. At the starred structured point this ratio is about $0.45$, while at one hundred times the cold exposure it is unity within numerical precision. Together these diagnostics isolate the intercycle-memory contribution while leaving the one-cycle statistics unchanged in the boundary-randomization test.

For the representative point in Fig.~\ref{fig:structured-faq}, the structured protocol has $\operatorname{Var}(w_1)\simeq2.2475$ and $D_W^{(\infty)}\simeq1.0109$, giving $\mathcal M_W\simeq0.450$. The corresponding equal-exposure flat reference has $\operatorname{Var}(w_1)\simeq1.9696$, $D_W^{(\infty)}\simeq1.3097$, and $\mathcal M_W\simeq0.665$. When the flat contact is retimed at fixed hot (cold) exposure allocation to match the structured mean work, its total exposure becomes $U_{\mathrm{flat}}\simeq0.09842$; then $\Pi_W^{(1)}\simeq0.00470$, $\Pi_W^{(\infty)}\simeq0.00656$, and $\mathcal M_W\simeq0.716$. Direct summation of the stationary lag covariances agrees with the resolvent-based asymptotic variance in all three cases.

For the representative microscopic rate $\gamma=4\times10^{-5}$ used in the resolved-channel assessment, the smallest frequency-resolution margins remain of order $13$ or larger at the starred point. The bath-memory product is about $5\times10^{-2}$, well below unity. At this coupling the point lies comfortably within the resolved weak-coupling regime used in the calculation.

The work-control comparison introduced in Sec.~\ref{sec:mesoscopic-controls} uses the general increment in Eq.~\eqref{eq:general-work-control}. A joint $(\omega,\lambda)$ stroke is compared with the frequency-only reference only when the excursion of every thermally active transition remains no larger than that of the frequency-only reference. The reservoir-control comparison fixes the per-contact spectral budget $B_r$ in Eq.~\eqref{eq:spectral-budget}. These constraints define the control classes without treating cancellation between $\Delta\omega$ and $\Delta\lambda$, or an increase of integrated reservoir strength, as a free resource.

\section{Mesoscopic boundary and control-comparison support}
\label{app:mesoscopic-boundary-support}

This appendix collects the finite-size validity conditions and the numerical control comparison underlying Sec.~\ref{sec:mesoscopic-controls}. The first part specifies the domain of the resolved-channel interpretation in Fig.~\ref{fig:mesoscopic-window}; the control comparison is reported separately in Table~\ref{tab:control-hierarchy-ratios}.

\subsection{Finite-size boundaries}

\begin{figure}[t]
\centering
\includegraphics[width=0.92\columnwidth]{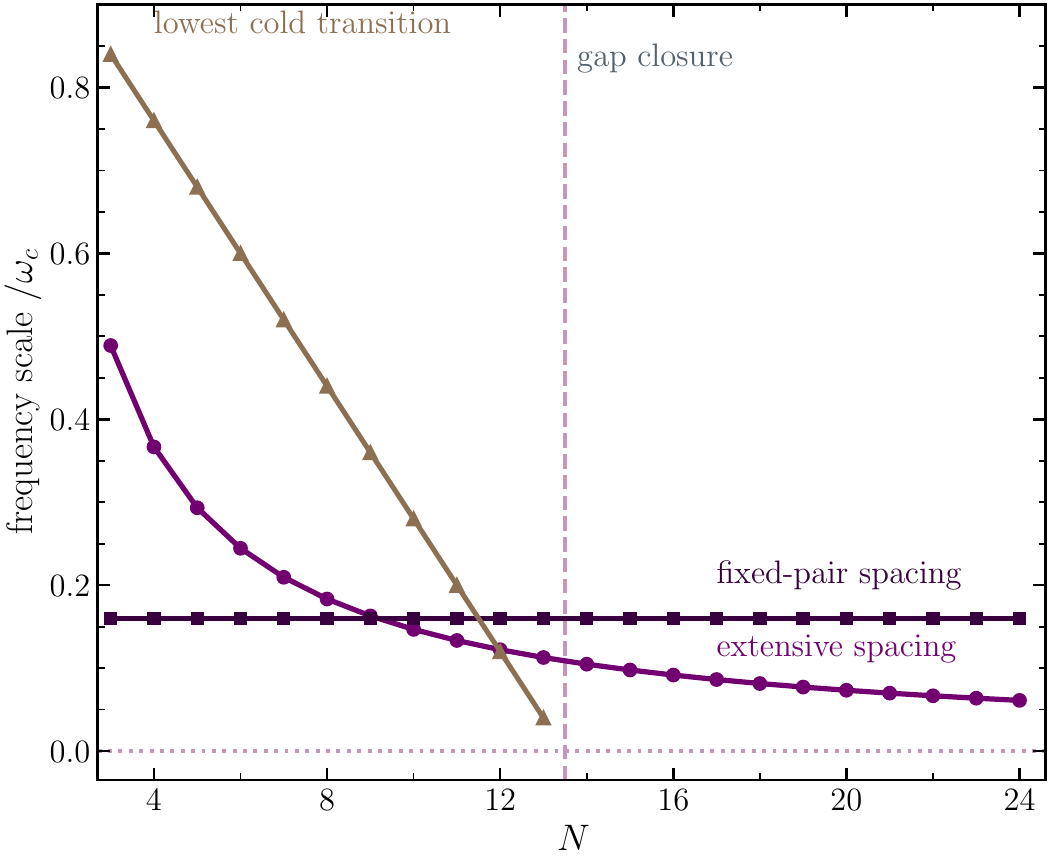}
\caption{\label{fig:mesoscopic-geometry-appendix} Technical finite-size spectral geometry.
The extensive continuation has neighboring-transition spacing $2\lambda_N=22/(15N)$, whereas the fixed-pair continuation keeps $2\lambda=0.16$. For the fixed-pair branch the lowest cold transition, $\Omega_{c,1}=1-0.08(N-1)$, softens with size and reaches its formal closure at $N=13.5$ (vertical guide); the $N=12$ operating point therefore remains on the positive-gap side. All spectral scales are shown in units of $\omega_c$.}
\end{figure}

The two size continuations lose their clean spectral interpretation for different reasons. For the extensive continuation $\lambda_N=(11/15)/N$, the neighboring-edge spacing contracts as $1/N$. At the reference microscopic coupling the conservative resolved-channel hierarchy remains clean only through about $N=12$; extending the same frequency-resolved Davies interpretation to larger $N$ requires weaker coupling and correspondingly longer physical contact times. For the fixed-pair continuation $\lambda=0.08$, the neighboring-edge spacing remains finite, but the lowest cold transition softens as the ladder length grows. Figure~\ref{fig:mesoscopic-geometry-appendix} summarizes the spectral boundaries underlying the size comparison, while the corresponding equal-output performance is shown in Fig.~\ref{fig:mesoscopic-window}.

\FloatBarrier

\subsection{Locally matched asymptotic routing residual}

The locally rate-matched asymptotic comparison isolates genuine spectral dispersion from the local kinetic amplitude. Fits over $N\ge 2048$ give the powers in Table~\ref{tab:asymptotic-routing-exponents}; smaller sizes are retained as preasymptotic points. Repeating the calculation with active low-excitation cutoffs $K_{\max}=80,100,120$ leaves the fitted scaling unchanged, while the stationary first law and the mean obtained from the tilted map agree with their direct evaluations.

\begin{table}[t]
\centering
\caption{Asymptotic shaped-minus-local scaling after local-rate matching. Fits use $N\ge2048$.}
\label{tab:asymptotic-routing-exponents}
\begin{tabular}{lcc}
\toprule
Observable & Expected power & Fitted power \\
\midrule
$|\Delta W|$ & $-1$ & $-1.098$ \\
$|\Delta\operatorname{Var}(W)|$ & $-1$ & $-0.988$ \\
$|\Delta D_W^{(\infty)}|$ & $-1$ & $-0.987$ \\
$|\Delta P_-|$ & $-1$ & $-0.992$ \\
$|\Delta\eta|$ & $-2$ & $-2.036$ \\
\bottomrule
\end{tabular}
\end{table}

Thus, after matching the local kinetic amplitude, the genuine multi-edge routing residual collapses as the transition window narrows. A budget-matched comparison can retain an $O(1)$ local kinetic offset because an integral spectral constraint does not force equality of the local pole amplitude, whereas the multi-edge routing residual vanishes with the narrowing transition window.

\subsection{Resolution--time tradeoff}

In the occupied low-excitation sector,
\begin{equation}
\Gamma_{\rm occ}\sim\gamma_N N,
\qquad
\Delta\Omega\sim N^{-1}.
\end{equation}
Maintaining an $N$-independent occupied-sector resolution ratio therefore requires approximately
\begin{equation}
\gamma_N=O(N^{-2}).
\end{equation}
Keeping the finite-contact map nontrivial then implies
\begin{equation}
\tau_N=\Omega(N),
\qquad
P_{\rm phys}=O(N^{-1})
\end{equation}
at best for this resolution-preserving sequence. Global resolution of the full Dicke ladder is more restrictive because the largest collective factors scale as $O(N^2)$. These relations delimit the individually resolved-edge description. Partially resolved or deliberately co-scaled large-$N$ reservoirs define a different generator regime.

In the reservoir-control comparison, the integrated spectral response of each thermal contact is held fixed,
\begin{equation}
B_r=\int_{\Omega_{r,-}}^{\Omega_{r,+}}
f_r(\Omega)\,d\Omega,
\label{eq:spectral-budget}
\end{equation}
where $\Omega_{r,-}$ and $\Omega_{r,+}$ delimit the angular-frequency band used to normalize contact $r$. Under this constraint, the spectral center and width can vary without changing the total spectral weight available within that band.

The control comparison uses the frequency/shared protocol as the baseline. Table~\ref{tab:control-hierarchy-ratios} gives the resulting gain factors. Each protocol class is first optimized for mean work under the active-edge work-excursion constraint and equal per-contact spectral budgets; the long-time precision is then evaluated at that same maximum-work point.

\begin{table}[t]
\centering
\caption{Control gains relative to the frequency-only/shared-spectrum baseline. Entries are $(\mathcal C_W,\mathcal C_\Pi)$ with $\mathcal C_X=X/X_{\rm base}$; long-time precision is evaluated at each protocol class's maximum-work point.}
\label{tab:control-hierarchy-ratios}
\scriptsize
\setlength{\tabcolsep}{2.2pt}
\begin{tabular}{@{}cccccc@{}}
\toprule
Family & $N$ & Base & \shortstack{joint\\$(\omega,\lambda)$} & \shortstack{indep.\\spectra} & \shortstack{both\\controls} \\
\midrule
extensive & 4 & $(1,1)$ & $(1.007,1.122)$ & $(1.033,0.984)$ & $(1.094,1.255)$ \\
extensive & 8 & $(1,1)$ & $(1.103,1.284)$ & $(1.051,0.976)$ & $(1.129,1.391)$ \\
extensive & 12 & $(1,1)$ & $(1.177,1.260)$ & $(1.037,0.879)$ & $(1.179,1.288)$ \\
fixed pair & 4 & $(1,1)$ & $(1.000,1.003)$ & $(1.080,1.075)$ & $(1.115,1.207)$ \\
fixed pair & 8 & $(1,1)$ & $(1.097,1.277)$ & $(1.102,0.938)$ & $(1.189,1.039)$ \\
fixed pair & 12 & $(1,1)$ & $(1.220,1.719)$ & $(1.005,1.095)$ & $(1.220,1.831)$ \\
\bottomrule
\end{tabular}
\end{table}

\FloatBarrier
\bibliography{references}

@article{Dicke1954,
  author        = {Dicke, R. H.},
  title         = {Coherence in Spontaneous Radiation Processes},
  journal       = {Physical Review},
  volume        = {93},
  pages         = {99--110},
  year          = {1954},
  doi           = {10.1103/PhysRev.93.99}
}

@article{Jaseem2023,
  author        = {Jaseem, Noufal and Vinjanampathy, Sai and Mukherjee, Victor},
  title         = {Quadratic enhancement in the reliability of collective quantum engines},
  journal       = {Physical Review A},
  volume        = {107},
  pages         = {L040202},
  year          = {2023},
  doi           = {10.1103/PhysRevA.107.L040202}
}

@misc{GarcezEtAlCollective2026,
  author        = {Garcez, Cler T. and Damas, Gabriella G. and de Almeida, Norton G. and de Moraes Neto, G. D.},
  title         = {Collective thermalization, work reliability, and resource bounds in a population-inverted {Dicke} {Otto} engine},
  year          = {2026},
  eprint        = {2609.30132},
  archivePrefix = {arXiv},
  primaryClass  = {quant-ph},
  doi           = {10.48550/arXiv.2609.30132},
  note          = {arXiv:2609.30132}
}

@article{Quan2005,
  author        = {Quan, H. T. and Zhang, P. and Sun, C. P.},
  title         = {Quantum heat engine with multilevel quantum systems},
  journal       = {Physical Review E},
  volume        = {72},
  pages         = {056110},
  year          = {2005},
  doi           = {10.1103/PhysRevE.72.056110}
}

@article{Quan2007,
  author        = {Quan, H. T. and Liu, Yu-xi and Sun, C. P. and Nori, Franco},
  title         = {Quantum thermodynamic cycles and quantum heat engines},
  journal       = {Physical Review E},
  volume        = {76},
  pages         = {031105},
  year          = {2007},
  doi           = {10.1103/PhysRevE.76.031105}
}

@article{RezekKosloff2006,
  author        = {Rezek, Yair and Kosloff, Ronnie},
  title         = {Irreversible Performance of a Quantum Harmonic Heat Engine},
  journal       = {New Journal of Physics},
  volume        = {8},
  pages         = {83},
  year          = {2006},
  doi           = {10.1088/1367-2630/8/5/083}
}

@article{Davies1974,
  author        = {Davies, E. B.},
  title         = {Markovian master equations},
  journal       = {Communications in Mathematical Physics},
  volume        = {39},
  pages         = {91--110},
  year          = {1974},
  doi           = {10.1007/BF01608389}
}

@article{GKS1976,
  author        = {Gorini, Vittorio and Kossakowski, Andrzej and Sudarshan, E. C. G.},
  title         = {Completely positive dynamical semigroups of {N}-level systems},
  journal       = {Journal of Mathematical Physics},
  volume        = {17},
  pages         = {821--825},
  year          = {1976},
  doi           = {10.1063/1.522979}
}

@article{Lindblad1976,
  author        = {Lindblad, G{\"o}ran},
  title         = {On the generators of quantum dynamical semigroups},
  journal       = {Communications in Mathematical Physics},
  volume        = {48},
  pages         = {119--130},
  year          = {1976},
  doi           = {10.1007/BF01608499}
}

@book{BreuerPetruccione2002,
  author        = {Breuer, Heinz-Peter and Petruccione, Francesco},
  title         = {The Theory of Open Quantum Systems},
  publisher     = {Oxford University Press},
  address       = {Oxford},
  year          = {2002},
  doi           = {10.1093/acprof:oso/9780199213900.001.0001}
}

@article{SchallerBrandes2008,
  author        = {Schaller, Gernot and Brandes, Tobias},
  title         = {Preservation of positivity by dynamical coarse graining},
  journal       = {Physical Review A},
  volume        = {78},
  pages         = {022106},
  year          = {2008},
  doi           = {10.1103/PhysRevA.78.022106}
}

@article{Trushechkin2021,
  author        = {Trushechkin, Anton},
  title         = {Unified {Gorini--Kossakowski--Lindblad--Sudarshan} quantum master equation beyond the secular approximation},
  journal       = {Physical Review A},
  volume        = {103},
  pages         = {062226},
  year          = {2021},
  doi           = {10.1103/PhysRevA.103.062226}
}

@article{TalknerLutzHanggi2007,
  author        = {Talkner, Peter and Lutz, Eric and H{\"a}nggi, Peter},
  title         = {Fluctuation theorems: Work is not an observable},
  journal       = {Physical Review E},
  volume        = {75},
  pages         = {050102(R)},
  year          = {2007},
  doi           = {10.1103/PhysRevE.75.050102}
}

@article{EspositoHarbolaMukamel2009,
  author        = {Esposito, Massimiliano and Harbola, Upendra and Mukamel, Shaul},
  title         = {Nonequilibrium fluctuations, fluctuation theorems, and counting statistics in quantum systems},
  journal       = {Reviews of Modern Physics},
  volume        = {81},
  pages         = {1665--1702},
  year          = {2009},
  doi           = {10.1103/RevModPhys.81.1665}
}

@article{CampisiHanggiTalkner2011,
  author        = {Campisi, Michele and H{\"a}nggi, Peter and Talkner, Peter},
  title         = {Colloquium: Quantum fluctuation relations: Foundations and applications},
  journal       = {Reviews of Modern Physics},
  volume        = {83},
  pages         = {771--791},
  year          = {2011},
  doi           = {10.1103/RevModPhys.83.771}
}

@article{MohantaAgarwalla2023,
  author        = {Mohanta, Sandipan and Agarwalla, Bijay Kumar},
  title         = {Full statistics of nonequilibrium heat and work for many-body quantum {Otto} engines and universal bounds: A nonequilibrium Green's function approach},
  journal       = {Physical Review E},
  volume        = {108},
  pages         = {064127},
  year          = {2023},
  doi           = {10.1103/PhysRevE.108.064127}
}

@article{Watanabe2017,
  author        = {Watanabe, Gentaro and Venkatesh, B. Prasanna and Talkner, Peter and del Campo, Adolfo},
  title         = {Quantum Performance of Thermal Machines over Many Cycles},
  journal       = {Physical Review Letters},
  volume        = {118},
  pages         = {050601},
  year          = {2017},
  doi           = {10.1103/PhysRevLett.118.050601}
}

@article{Jones2013TunableEnvironment,
  author        = {Jones, P. J. and Huhtam{\"a}ki, J. A. M. and Salmilehto, J. and Tan, K. Y. and M{\"o}tt{\"o}nen, M.},
  title         = {Tunable electromagnetic environment for superconducting quantum bits},
  journal       = {Scientific Reports},
  volume        = {3},
  pages         = {1987},
  year          = {2013},
  doi           = {10.1038/srep01987}
}

@article{Wang2019CavityAttenuator,
  author        = {Wang, Z. and Shankar, S. and Minev, Z. K. and Campagne-Ibarcq, P. and Narla, A. and Devoret, M. H.},
  title         = {Cavity Attenuators for Superconducting Qubits},
  journal       = {Physical Review Applied},
  volume        = {11},
  pages         = {014031},
  year          = {2019},
  doi           = {10.1103/PhysRevApplied.11.014031}
}

@article{Scigliuzzo2020MicrowaveThermometry,
  author        = {Scigliuzzo, M. and Bengtsson, A. and Besse, J.-C. and Wallraff, A. and Delsing, P. and Gasparinetti, S.},
  title         = {Primary Thermometry of Propagating Microwaves in the Quantum Regime},
  journal       = {Physical Review X},
  volume        = {10},
  pages         = {041054},
  year          = {2020},
  doi           = {10.1103/PhysRevX.10.041054}
}

@article{Yan2023BroadbandPurcell,
  author        = {Yan, H. and Wu, X. and Lingenfelter, A. and Joshi, Y. J. and Andersson, G. and Conner, C. R. and Chou, M.-H. and Grebel, J. and Miller, J. M. and Povey, R. G. and Qiao, H. and Clerk, A. A. and Cleland, A. N.},
  title         = {Broadband bandpass {Purcell} filter for circuit quantum electrodynamics},
  journal       = {Applied Physics Letters},
  volume        = {123},
  pages         = {134001},
  year          = {2023},
  doi           = {10.1063/5.0161893}
}

@article{GrossHaroche1982,
  author        = {Gross, M. and Haroche, S.},
  title         = {Superradiance: An essay on the theory of collective spontaneous emission},
  journal       = {Physics Reports},
  volume        = {93},
  number        = {5},
  pages         = {301--396},
  year          = {1982},
  doi           = {10.1016/0370-1573(82)90102-8}
}

@article{Kloc2019,
  author        = {Kloc, Michal and Cejnar, Pavel and Schaller, Gernot},
  title         = {Collective performance of a finite-time quantum {Otto} cycle},
  journal       = {Physical Review E},
  volume        = {100},
  pages         = {042126},
  year          = {2019},
  doi           = {10.1103/PhysRevE.100.042126}
}

@article{Kamimura2022,
  author        = {Kamimura, Shunsuke and Hakoshima, Hideaki and Matsuzaki, Yuichiro and Yoshida, Kyo and Tokura, Yasuhiro},
  title         = {Quantum-Enhanced Heat Engine Based on Superabsorption},
  journal       = {Physical Review Letters},
  volume        = {128},
  pages         = {180602},
  year          = {2022},
  doi           = {10.1103/PhysRevLett.128.180602}
}

@article{Souza2022,
  author        = {da Silva Souza, Leonardo and Manzano, Gonzalo and Fazio, Rosario and Iemini, Fernando},
  title         = {Collective effects on the performance and stability of quantum heat engines},
  journal       = {Physical Review E},
  volume        = {106},
  pages         = {014143},
  year          = {2022},
  doi           = {10.1103/PhysRevE.106.014143}
}

@article{Higgins2014,
  author        = {Higgins, K. D. B. and Benjamin, S. C. and Stace, T. M. and Milburn, G. J. and Lovett, B. W. and Gauger, E. M.},
  title         = {Superabsorption of light via quantum engineering},
  journal       = {Nature Communications},
  volume        = {5},
  pages         = {4705},
  year          = {2014},
  doi           = {10.1038/ncomms5705}
}

@article{Naseem2020,
  author        = {Naseem, M. Tahir and Misra, Avijit and M{\"u}stecapl{\i}o{\u{g}}lu, {\"O}zg{\"u}r E. and Kurizki, Gershon},
  title         = {Minimal quantum heat manager boosted by bath spectral filtering},
  journal       = {Physical Review Research},
  volume        = {2},
  pages         = {033285},
  year          = {2020},
  doi           = {10.1103/PhysRevResearch.2.033285}
}

@article{GelbwaserKlimovsky2013,
  author        = {Gelbwaser-Klimovsky, David and Alicki, Robert and Kurizki, Gershon},
  title         = {Minimal universal quantum heat machine},
  journal       = {Physical Review E},
  volume        = {87},
  pages         = {012140},
  year          = {2013},
  doi           = {10.1103/PhysRevE.87.012140}
}

@article{DenzlerLutz2020,
  author        = {Denzler, Tobias and Lutz, Eric},
  title         = {Efficiency fluctuations of a quantum heat engine},
  journal       = {Physical Review Research},
  volume        = {2},
  pages         = {032062(R)},
  year          = {2020},
  doi           = {10.1103/PhysRevResearch.2.032062}
}

@article{DenzlerEtAl2024,
  author        = {Denzler, Tobias and Santos, Jonas F. G. and Lutz, Eric and Serra, Roberto M.},
  title         = {Nonequilibrium fluctuations of a quantum heat engine},
  journal       = {Quantum Science and Technology},
  volume        = {9},
  pages         = {045017},
  year          = {2024},
  doi           = {10.1088/2058-9565/ad6287}
}

@article{SaryalAgarwalla2021,
  author        = {Saryal, Sushant and Agarwalla, Bijay Kumar},
  title         = {Bounds on fluctuations for finite-time quantum {Otto} cycle},
  journal       = {Physical Review E},
  volume        = {103},
  pages         = {L060103},
  year          = {2021},
  doi           = {10.1103/PhysRevE.103.L060103}
}

@article{Bouton2021,
  author        = {Bouton, Quentin and Nettersheim, Jens and Burgardt, Sabrina and Adam, Daniel and Lutz, Eric and Widera, Artur},
  title         = {A quantum heat engine driven by atomic collisions},
  journal       = {Nature Communications},
  volume        = {12},
  pages         = {2063},
  year          = {2021},
  doi           = {10.1038/s41467-021-22222-z}
}

@article{HardalMustecaplioglu2015,
  author        = {Hardal, Ali {\"U}. C. and M{\"u}stecapl{\i}o{\u{g}}lu, {\"O}zg{\"u}r E.},
  title         = {Superradiant Quantum Heat Engine},
  journal       = {Scientific Reports},
  volume        = {5},
  pages         = {12953},
  year          = {2015},
  doi           = {10.1038/srep12953}
}

@article{NiedenzuKurizki2018,
  author        = {Niedenzu, Wolfgang and Kurizki, Gershon},
  title         = {Cooperative many-body enhancement of quantum thermal machine power},
  journal       = {New Journal of Physics},
  volume        = {20},
  pages         = {113038},
  year          = {2018},
  doi           = {10.1088/1367-2630/aaed55}
}

@article{Uusnakki2026,
  author        = {Uusn{\"a}kki, Tuomas and M{\"o}rstedt, Timm and Teixeira, Wallace and Rasola, Miika and M{\"o}tt{\"o}nen, Mikko},
  title         = {Initial demonstration of a quantum heat engine based on dissipation-engineered superconducting circuits},
  journal       = {Nature Communications},
  volume        = {17},
  pages         = {6054},
  year          = {2026},
  doi           = {10.1038/s41467-026-72651-x}
}

@article{Aamir2025,
  author        = {Aamir, Mohammed Ali and Jamet Suria, Paul and Mar{\'i}n Guzm{\'a}n, Jos{\'e} Antonio and Castillo-Moreno, Claudia and Epstein, Jeffrey M. and Yunger Halpern, Nicole and Gasparinetti, Simone},
  title         = {Thermally driven quantum refrigerator autonomously resets a superconducting qubit},
  journal       = {Nature Physics},
  volume        = {21},
  pages         = {318--323},
  year          = {2025},
  doi           = {10.1038/s41567-024-02708-5}
}

@article{Song2019,
  author        = {Song, Chao and Xu, Kai and Li, Hekang and Zhang, Yuran and Zhang, Xu and Liu, Wuxin and Guo, Qiujiang and Wang, Zhen and Ren, Wenhui and Hao, Jie and Feng, Hui and Fan, Heng and Zheng, Dongning and Wang, Dawei and Wang, H. and Zhu, Shiyao},
  title         = {Generation of multicomponent atomic Schr{\"o}dinger cat states of up to 20 qubits},
  journal       = {Science},
  volume        = {365},
  pages         = {574--577},
  year          = {2019},
  doi           = {10.1126/science.aay0600}
}

@article{Wang2020Superradiance,
  author        = {Wang, Zhen and Li, Hekang and Feng, Wei and Song, Xiaohui and Song, Chao and Liu, Wuxin and Guo, Qiujiang and Zhang, Xu and Dong, Hang and Zheng, Dongning and Wang, H. and Wang, Da-Wei},
  title         = {Controllable Switching between Superradiant and Subradiant States in a 10-qubit Superconducting Circuit},
  journal       = {Physical Review Letters},
  volume        = {124},
  pages         = {013601},
  year          = {2020},
  doi           = {10.1103/PhysRevLett.124.013601}
}

@article{AlickiGelbwaserKlimovsky2015,
  author        = {Alicki, Robert and Gelbwaser-Klimovsky, David},
  title         = {Non-equilibrium quantum heat machines},
  journal       = {New Journal of Physics},
  volume        = {17},
  pages         = {115012},
  year          = {2015},
  doi           = {10.1088/1367-2630/17/11/115012}
}

@article{Blais2021,
  author        = {Blais, Alexandre and Grimsmo, Arne L. and Girvin, S. M. and Wallraff, Andreas},
  title         = {Circuit quantum electrodynamics},
  journal       = {Reviews of Modern Physics},
  volume        = {93},
  pages         = {025005},
  year          = {2021},
  doi           = {10.1103/RevModPhys.93.025005}
}

@Article{Prataviera2026,
AUTHOR = {Prataviera, Gilberto Aparecido and de Oliveira, Marcos César},
TITLE = {Finite-Bandwidth Protection of a Three-Level Quantum Heat Engine Against Parasitic Heat Leaks},
JOURNAL = {Entropy},
VOLUME = {28},
YEAR = {2026},
NUMBER = {10},
ARTICLE-NUMBER = {1081},
URL = {https://www.mdpi.com/1099-4300/28/10/1081},
PubMedID = {42940803},
ISSN = {1099-4300},
DOI = {10.3390/e28101081}
}

@article{Xu2024,
  title = {Universal quantum Otto heat machine based on the Dicke model},
  author = {Xu, He-Guang and Jin, Jiasen and Neto, G. D. M. and de Almeida, Norton G.},
  journal = {Phys. Rev. E},
  volume = {109},
  issue = {1},
  pages = {014122},
  numpages = {14},
  year = {2024},
  month = {Jan},
  publisher = {American Physical Society},
  doi = {10.1103/PhysRevE.109.014122},
  url = {https://link.aps.org/doi/10.1103/PhysRevE.109.014122}
}

@misc{Burgardt2026,
      title={Enhancing the power of a quantum heat engine via control of the system--reservoir coupling}, 
      author={Sabrina Burgardt and Julian Feß and Silvia Hiebel and Eric Lutz and Artur Widera},
      year={2026},
      eprint={2608.12055},
      archivePrefix={arXiv},
      primaryClass={quant-ph},
      url={https://arxiv.org/abs/2608.12055}, 
}

@article{Xu2022,
  title = {Correlation-enhanced stability of microscopic cyclic heat engines},
  author = {Xu, Guo-Hua and Watanabe, Gentaro},
  journal = {Phys. Rev. Res.},
  volume = {4},
  issue = {3},
  pages = {L032017},
  numpages = {7},
  year = {2022},
  month = {Aug},
  publisher = {American Physical Society},
  doi = {10.1103/PhysRevResearch.4.L032017},
  url = {https://link.aps.org/doi/10.1103/PhysRevResearch.4.L032017}
}

@article{Boubakour2023,
  title = {Interaction-enhanced quantum heat engine},
  author = {Boubakour, Mohamed and Fogarty, Thom{\'a}s and Busch, Thomas},
  journal = {Phys. Rev. Res.},
  volume = {5},
  issue = {1},
  pages = {013088},
  numpages = {10},
  year = {2023},
  month = {Feb},
  publisher = {American Physical Society},
  doi = {10.1103/PhysRevResearch.5.013088},
  url = {https://link.aps.org/doi/10.1103/PhysRevResearch.5.013088}
}

@article{Anka2024,
  title = {Work and efficiency fluctuations in a quantum Otto cycle with idle levels},
  author = {Anka, Maron F. and de Oliveira, Thiago R. and Jonathan, Daniel},
  journal = {Phys. Rev. E},
  volume = {109},
  issue = {6},
  pages = {064129},
  numpages = {12},
  year = {2024},
  month = {Jun},
  publisher = {American Physical Society},
  doi = {10.1103/PhysRevE.109.064129},
  url = {https://link.aps.org/doi/10.1103/PhysRevE.109.064129}
}

@article{Kloc2021,
  title = {Superradiant Many-Qubit Absorption Refrigerator},
  author = {Kloc, Michal and Meier, Kurt and Hadjikyriakos, Kimon and Schaller, Gernot},
  journal = {Phys. Rev. Appl.},
  volume = {16},
  issue = {4},
  pages = {044061},
  numpages = {15},
  year = {2021},
  month = {Oct},
  publisher = {American Physical Society},
  doi = {10.1103/PhysRevApplied.16.044061},
  url = {https://link.aps.org/doi/10.1103/PhysRevApplied.16.044061}
}

@article{Andolina2024,
  title = {Dicke superradiant heat current enhancement in circuit quantum electrodynamics},
  author = {Andolina, Gian Marcello and Erdman, Paolo Andrea and No{\'e}, Frank and Pekola, Jukka and Schir{\`o}, Marco},
  journal = {Phys. Rev. Res.},
  volume = {6},
  issue = {4},
  pages = {043128},
  numpages = {17},
  year = {2024},
  month = {Nov},
  publisher = {American Physical Society},
  doi = {10.1103/PhysRevResearch.6.043128},
  url = {https://link.aps.org/doi/10.1103/PhysRevResearch.6.043128}
}

@article{Boeyens2025,
doi = {10.1088/1367-2630/ada796},
url = {https://doi.org/10.1088/1367-2630/ada796},
year = {2025},
month = {jan},
publisher = {IOP Publishing},
volume = {27},
number = {1},
pages = {013021},
author = {Boeyens, Julia and Yadin, Benjamin and Nimmrichter, Stefan},
title = {How exchange symmetry impacts performance of collective quantum heat engines},
journal = {New Journal of Physics}
}
\end{document}